\documentclass[fleqn,usenatbib]{mnras}

\usepackage{newtxtext,newtxmath}
\usepackage[T1]{fontenc}

\DeclareRobustCommand{\VAN}[3]{#2}
\let\VANthebibliography\thebibliography
\def\thebibliography{\DeclareRobustCommand{\VAN}[3]{##3}\VANthebibliography}

\usepackage{graphicx}	
\usepackage{amsmath}	
\usepackage{pdflscape}	

\usepackage{siunitx}
\usepackage{dcolumn}
\newcolumntype{d}[1]{D{.}{.}{#1}}

\title[Spectropolarimetric behaviour of blazars]{The optical spectropolarimetric behaviour of a selection of high-energy blazars}

\author[J. Barnard et al.]{
Joleen Barnard,$^{1}$\thanks{E-mail: elsj@ufs.ac.za}
B. van Soelen,$^{1}$
S. Acharya,$^{2}$
M. B\"{o}ttcher,$^{3}$
R. J. Britto,$^{1}$
J. Cooper,$^{1}$
D. A. H. Buckley,$^{1,4}$
\newauthor
A. Martin-Carrillo,$^{5}$
B. Vaidya,$^{6}$
I. P. van der Westhuizen,$^{1}$
and M. Zacharias$^{7,3}$
\\
$^{1}$Department of Physics, University of the Free State, Bloemfontein, 9301, South Africa\\
$^{2}$Hamburger Sternwarte, Universität Hamburg, Gojenbergsweg 112, 21029 Hamburg, Germany\\
$^{3}$Centre for Space Research, North West University, Potchefstroom, 2520, South Africa\\
$^{4}$South African Astronomical Observatory, Observatory, Cape Town, 7925, South Africa\\
$^{5}$Space Science Group, School of Physics, University College Dublin, Dublin, Ireland\\
$^{6}$Discipline of Astronomy Astrophysics and Space Engineering, Indian Institute of Technology Indore, Indore, India\\
$^{7}$Landessternwarte, Universität Heidelberg, Königstuhl 12, 69117 Heidelberg, Germany
}

\date{Accepted XXX. Received YYY; in original form ZZZ}

\pubyear{2023}

\begin{document}
\label{firstpage}
\pagerange{\pageref{firstpage}--\pageref{lastpage}}
\maketitle

\begin{abstract}
At optical/ultraviolet energies, blazars display an underlying thermal (unpolarized) contribution from the accretion disc, torus and line emitting regions, diluting the polarized emission from the jet-component. Optical polarimetry can be used to disentangle the thermal and non-thermal components, and place constraints on the particle populations and acceleration mechanisms responsible for the non-thermal emission. We present the results of a linear optical spectropolarimetric observing campaign of 18 blazars (6 BLLs and 12 FSRQs) undertaken with the Southern African Large Telescope between 2016 and 2022. This was done to observe these systems during flaring states, as well as long term monitoring of PKS1510-089, AP Lib and PKS 1034-293. The observations traced the frequency dependence of the degree and angle of polarization, as well as changes in the spectral line strengths. We investigated possible correlations between the polarization and other observed characteristics for the sources. While an indication of correlation was found between the frequency dependence and the average level of polarization for some sources, a correlation was not found for the population as a whole. These results highlight that continuous observations and in-depth modelling of polarization and its frequency dependence is required to obtain a more holistic view of TeV blazars.

\end{abstract}




\section{Introduction}
\label{sec:intro}

Active Galactic Nuclei (AGN) are the central active cores of some galaxies. These sources are believed to be powered by the accretion of material onto a central supermassive black hole \citep[SMBH;][]{2017FrASS...4...35P, 2017A&ARv..25....2P}. AGN are characterised by very high luminosities emitted across the entire electromagnetic spectrum that can, in many cases, outshine the total emission of the host galaxy itself \citep{1995PASP..107..803U, 2017SSRv..207....5R}. They also exhibit variability on timescales from minutes \citep[e.g. Mrk 501;][]{2007ApJ...669..862A} to years \citep[see e.g.][and references therein]{2017FrASS...4...35P}. AGN have an approximately axisymmetric structure, and the type of AGN observed depends largely on the viewing angle, rate of accretion onto the SMBH, and the presence/absence of a relativistic jet (i.e. jetted versus non-jetted AGN).

Blazars form part of the jetted class of AGN, with the relativistic jet closely aligned to the observer's line of sight. This viewing geometry results in blazars displaying highly variable, Doppler-boosted, non-thermal emission ranging from radio to $\gamma$-ray wavelengths. Blazars are subdivided into two different classes: BL Lac-type objects (BLLs), and Flat-Spectrum Radio Quasars (FSRQs). They are mainly classified based on their optical spectral features, where BLLs display a largely featureless continuum and/or weak, narrow absorption features with equivalent widths set at $|\rm W_\lambda| < 5$\,{\rm \AA}. FSRQs, however, have strong, broad emission features with $|\rm W_\lambda| > 5$\,{\rm \AA} \citep{1998MNRAS.299..433F, 2011ApJ...740...98M, 1995MNRAS.277.1477P, 1995ApJ...444..567P}. Blazars can also be classified based on the location of the synchrotron peak frequency ($\nu_{\rm sy}$) and are divided into low-synchrotron peaked sources (LSPs, or LBLs in the case of BLLs), intermediate-synchrotron peaked sources (ISPs, or IBLs), and high-synchrotron peaked sources (HSPs, or HBLs) \citep[see e.g.][]{1998MNRAS.299..433F, 2011ApJ...740...98M}. Furthermore, the flux variability, photon indices, and $\gamma$-ray dominance of blazars can also be used as a method of classification \citep{2010ApJ...710.1271A, 2010ApJ...716...30A}.

The bulk of the observed blazar emission originates from the jet, and the spectral energy distributions (SEDs) are characterised by two broad, non-thermal components \citep[e.g.][]{1998MNRAS.301..451G, 1999APh....11..159U, 2019Galax...7...20B}. The low-energy component spans the radio to optical/UV range, and can sometimes extend into the soft X-ray regime in the presence of different particle acceleration mechanisms \citep{1998A&A...333..452K}. For example, \citet{2023A&A...671A.161A} showed that the low energy component could extend till soft X-ray due to the presence of the shocks generated with the evolution of MHD instability. Further, particles could also accelerate up to high energy via magnetic reconnection, and that in turn affects the observed polarization properties \citep{2021MNRAS.501.2836B}. This component is believed to be dominated by leptonic synchrotron emission from relativistic electrons in the jet, which is intrinsically polarized. At infrared/optical/UV wavelengths, there is also an underlying thermal, unpolarized contribution from the accretion disc, dust torus, broad-line region (BLR), and host galaxy \citep{2019Galax...7...20B, 2019Galax...7...85Z}. Thus, at optical wavelengths, the observed emission is a superposition of polarized (non-thermal) and unpolarized (thermal) light. Since the non-thermal jet emission is highly variable, the contribution of the non-thermal emission fluctuates with time, leading to a variable degree of polarization observed in blazars \citep[e.g.][]{2017Galax...5...52B,2022ApJ...925..139S}.

The high-energy component (X-ray to $\gamma$-ray energies) can be reproduced by two different types of models, namely leptonic or hadronic models. In the leptonic scenario, the high-energy emission is assumed to be produced by the inverse-Compton (IC) scattering of low-energy photons by the same population of relativistic electrons producing the low-energy component. This can happen either via synchrotron self-Compton (SCC) or external Compton (EC) \citep{2013ApJ...768...54B, 1992ApJ...397L...5M}. The hadronic scenario assumes that protons and/or nuclei are sufficiently accelerated to produce the high-energy component via proton-synchrotron emission and photo-pion production \citep{1993A&A...269...67M, 2019Galax...7...85Z}. While the leptonic scenario is successful in reproducing the spectral energy distributions and variability patterns of many blazars,
it fails to explain the occurrence of neutrino production and ``orphan'' flares (i.e. flares occurring only in one energy band), which hadronic models can successfully explain \citep{2016CRPhy..17..594D, 2019Galax...7...20B}. However, in some cases, both a leptonic and hadronic approach to modelling the SEDs of blazars are equally successful \citep[see, e.g.,][]{2013ApJ...768...54B}. Thus, constraining and/or disentangling the emission of blazars using SED modelling alone is often not possible.

Polarimetry can be used as a powerful diagnostic tool in blazar studies, as it provides a way to disentangle the thermal and non-thermal emission at optical wavelengths, constrain the jet's magnetic field, and determine the mass of the SMBH. It also aides in constraining the high-energy component of blazar emission in the leptonic scenario, as the same population of electrons is assumed to produce both the low and high-energy components \citep[see e.g.][]{2022ApJ...925..139S}. 

The polarization characteristics vary from blazar to blazar. Its frequency dependence at optical wavelengths depends on a multitude of factors, e.g. the redshift ($z$) of the source, the location of the synchrotron peak frequency ($\nu_{\rm sy}$), the ordering of the magnetic field ($F_{\rm B}$), etc. In addition to intrinsic characteristics, polarization properties can be modeled using geometrical frameworks such as twisted or helical jet models \citep[see e.g.][and references therein]{2023MNRAS.526.4502R}. The effects and strength of the accretion disc component should also be considered, as FSRQs generally exhibit more powerful accretion discs 
than BLLs. For example, in the model by \citet{2022ApJ...925..139S}, the authors found a decrease in the degree of linear polarization towards higher (optical/UV) frequencies for the FSRQ 4C+01.02 during both flaring and quiescent states. This was attributed to the dilution of the degree of polarization by the thermal accretion disc component.

However, an increase in the degree of polarization towards higher frequencies might be due to a shock in the blazar jet \citep{2018MNRAS.480.2872T}. Here, a rapidly decaying magnetic field component perpendicular to the direction of the shock's motion ($B_{\perp}$) is generated due to shock compression and instabilities in the vicinity of the shock front. Close to the shock region, $B_{\perp}$ is highly ordered, leading to higher degrees of polarization observed for the more energetic electrons (e.g. emitting at X-ray wavelengths). Further downstream along the jet, radiative cooling processes cause the electrons to cool rapidly, and the quasi-parallel magnetic field component becomes dominant over $B_{\perp}$. Therefore, lower-energy particles emitting synchrotron radiation at lower frequencies (optical/infrared), probing a larger volume behind the shock front, experience a less ordered magnetic field, and the observed polarization will be lower. This is illustrated in fig.~3 of \citet{2018MNRAS.480.2872T}.

Recently, large-scale data collection campaigns have been launched to investigate the nature of polarization in AGN in general and blazars in particular. Some of these polarimetry surveys include MOBPol,\footnote{\url{https://www.bu.edu/blazars/mobpol/mobpol.html}} the Steward Observatory's Ground-based Observational Support of the \textit{Fermi} Gamma-ray Space Telescope at the University of Arizona,\footnote{\url{http://james.as.arizona.edu/~psmith/Fermi/}}, the systematic study of $\gamma$-ray bright blazars using the Kanata telescope \citep{2016ApJ...833...77I}, as well as the RoboPol programme.\footnote{\url{https://robopol.physics.uoc.gr/}}

The RoboPol programme (2014 - 2017) is the leading effort in photo-polarimetry surveys, and aims at investigating statistical trends in a large sample of $\gamma$-ray bright blazars as it connects to the polarization signatures in their optical emission. The relation between $\nu_{\rm{sy}}$ and the radio-loudness of AGN, the variability and statistical distribution of the degree
of polarization in blazars, and the relation between polarization angle rotations and $\gamma$-ray flaring activity outline some of the RoboPol collaboration's findings \citep{2016MNRAS.463.3365A,2014MNRAS.442.1693P,2016A&A...596A..78H,2015MNRAS.453.1669B,2016MNRAS.457.2252B,2016MNRAS.462.1775B,2018MNRAS.474.1296B}. Note, however, that the photo-polarimetry data from RoboPol do not reveal the spectral characteristics of the polarization. Another blazar polarimetry survey was performed by the Steward observatory, who found a clear difference between the polarization of FSRQs/LSPs and ISPs/HSPs, as well as a preferential orientation of the polarization angle in BL Lacs which was not found for FSRQs \citep[see ][for a full discussion of their results]{2023MNRAS.523.4504O}.

In this paper, we present optical spectropolarimetric observations of a sample of eighteen blazars (six BLLs and twelve FSRQs), complemented by optical photometric and $\gamma$-ray observations. Long-term monitoring was undertaken for three sources (AP Lib, PKS 1034--293, PKS 1510--089), while the remaining sources were observed around periods of increased $\gamma$-ray activity. For this sample of blazars, we investigate the polarization characteristics in an attempt to a) trace the polarization as it transitions from a flaring to a quiescent state, b) track the long-term evolution of the polarization in some of the sources, and c) investigate the overall behaviour of the blazar population as a whole. This was done by investigating possible correlations between the polarization and other observed characteristics of the blazars. To measure these correlations, the Spearman's test was used. Throughout the text, a strong correlation/anti-correlation is defined has having $\rho > + 0.5$ or $\rho < -0.5$, respectively, and the correlation is significant when it  has a p-value $< 0.05$.

This paper is structured as follows: Section \ref{sec:obs_setup} presents the observational setups and data reduction processes; Section \ref{sec:results}  highlights some of the key observational results obtained; Section \ref{sec:discussion} presents a discussion of the observational results, highlighting some of the important statistical results; and lastly the conclusions are presented in Section \ref{sec:conclusions}. The summary of all the observations is given in Appendix~\ref{app:full_results}.

\section{Observations}
\label{sec:obs_setup}

This section gives an overview of the observational setups and data reduction processes used to obtain the optical spectropolarimetric, optical photometric, and $\gamma$-ray observations.

\subsection{Optical spectropolarimetry}
\label{subsec:phot}

All of the optical spectropolarimetric observations were taken with the Southern African Large Telescope \citep[SALT;][]{2006_SALT_Buckley}, using the Robert Stobie Spectrograph \citep[RSS][]{2003SPIE.4841.1463B,2003SPIE.4841.1634K}, located at the South African Astronomical Observatory (SAAO). All of the science observations were taken with the RSS in either longslit spectropolarimetry \textsc{linear} mode (with the $1/2$ waveplate at angles $0^{\circ}$, $45^{\circ}$, $22.5^{\circ}$, and $67.5^{\circ}$) or \textsc{linear-hi} mode (with additional observations taken with the $1/2$ waveplate at $11.25^{\circ}$, $56.25^{\circ}$, $33.75^{\circ}$, and $78.75^{\circ}$). The observations were taken using either the PG0300 or PG0900 diffraction grating at various grating angles and slit widths, depending on the source. Arc frames were taken immediately after the science cycle using an Ar, ThAr, Xe, or Ne arc lamp, depending on the spectrograph configuration. A total of sixty-eight (68) optical spectropolarimetry observations were obtained for the eighteen different sources, between 2016 and 2022. The different observational setups for all of the sources are summarised in Table~\ref{tab:obs_properties}, with the results obtained from these observations summarised in Table~\ref{tab:full_results} in Appendix \ref{app:full_results}. The errors quoted for the SALT data are statistical.

As we are observing extra-galactic sources, typically at high galactic latitudes, the effects of interstellar polarization (ISP) on the blazar's observed polarization level and its frequency dependence should be low. The typical and maximum expected levels of ISP depends on the extinction, ${\rm E(B-V)}$, along the line of sight to each target \citep{1975ApJ...196..261S, Smith04}; i.e.
\begin{equation}
    \rm{ISP_{max}} = 9 \times E(B - V),
    \label{eqn:ISP_max}
\end{equation}
and
\begin{equation}
    \rm{ISP_{typical}} = 3 \times E(B-V).
    \label{eqn:ISP_typical}
\end{equation}
Additionally, according to \citet{2005MNRAS.363.1241H}, the maximum polarization angle change, for a given ISP magnitude, can be determined by
\begin{equation}
    \Delta \rm{PA_{max}} = 0.5 \sin^{-1} \left( \frac{\rm{ISP}}{\Pi_{\rm{obs}}} \right),
    \label{eqn:DeltaPA_max}
\end{equation}
where ISP is the typical magnitude of interstellar polarization, and $\Pi_{\rm{obs}}$ is the observed degree of polarization from the target. Based on the Galactic Extinction, the maximum ISP for our targets is $<1$ per cent (Table~\ref{tab:ISP_properties}). The observations have, therefore, not been corrected for the ISP.

Additionally, most of the observations were taken such that a comparison star fell elsewhere on the slit. The average degree of polarization for the comparison stars was found to be below 3 per cent. The frequency dependence of the comparison stars' polarization was also investigated and the slopes were found to be very flat. This suggests that the ISP does not introduce a significant wavelength dependence, and any effect from the ISP falls well within the systematic uncertainty of the observations, which we take as 3 per cent. However, we caution that uncertainties in the polarization angle may be significant when the magnitude of the ISP and observed degree of polarization of the target becomes comparable (i.e. when the target's degree of polarization is on the order of $\sim$3\%).

\begin{table*}
\begin{center}
\small
\caption{The SALT RSS spectropolarimetry observational setups for all sources. The observation dates, grating, grating angle, arc lamp used, slit width used, resulting resolving power, and exposure time of all of the blazars observed are listed.
Note that the exposure time is given as the exposure time per $1/2$ waveplate angle, multiplied by the number of $1/2$ waveplate angles used, with four corresponding to \textsc{linear} mode, and eight corresponding to \textsc{linear-hi} mode.}
\label{tab:obs_properties}
\begin{tabular}{lccd{2.3}cd{1.2}cc}
\hline
Target & Obs. Dates & Grating & \multicolumn{1}{c}{Grating} & Arc  & \multicolumn{1}{c}{Slit}      & Resolving & Exposure   \\
       &            &         & \multicolumn{1}{c}{Angle ($^{\circ}$)} & Lamp & \multicolumn{1}{c}{Width (")} &  Power    &  Time (s) \\
\hline
PKS 0426--380      & 2017-01-21 & PG0300 & 5.375 &  Ar  & 1.5 & 160 -- 530 & 600$\times$4 \cr
                   & 2017-02-20 &        &        &      &              \cr
PKS 0447--439      & 2017-02-21 & PG0900 & 19.625 &  Ne  & 1.5 & 1150 -- 1515 &  120$\times$4 \cr
                   &            &        & 12.500 & ThAr & 1.5 & 640 -- 1010 &              \cr 
TXS 0506+056       & 2017-10-15 & PG0900 & 14.000 &  Ar  & 1.5 & 750 -- 1100 & 600$\times$4 \cr
                   & 2017-10-21 &        & 12.875 & ThAr & & &               \cr
PKS 0537--441      & 2019-01-14 & PG0900 & 14.375 &  Ar  & 1.5 & 770 -- 1140 & 650$\times$4 \cr
                   & 2019-03-05 &        &        &      &              \cr
PKS 1454--354      & 2021-06-13 & PG0300 & 5.375 &  Ar  & 1.5 & 160 -- 530 & 600$\times$4 \cr
AP Lib             & 2020-05-14 - & PG0300 & 5.375 &  Ar  & 1.5 & 160 -- 530 & 30$\times$8  \cr
                   & 2021-07-31   &        &        &      &              \cr
PKS 0035--252      & 2018-07-21 & PG0300 & 5.375 &  Ar  & 1.5 & 160 -- 530 & 625$\times$4 \cr
PKS 0131--522      & 2017-11-19 & PG0300 & 5.375 &  Ar  & 1.5 & 160 -- 530 & 800$\times$4 \cr
                   & 2017-11-22 &        &        &      &              \cr
PKS 0208--512      & 2019-12-05 & PG0300 & 5.750 &  Ar  & 1.5 & 190 -- 560 &  300$\times$8 \cr
                   & 2019-12-19 &        &        &      &              \cr
4FGL J0231.2--4745 & 2019-10-23 & PG0900 & 15.500 &  Xe  & 1.5 & 850 -- 1220 &  274$\times$8 \cr
                   & 2019-10-29 &        &        &      &              \cr
PKS 0346--279      & 2018-02-09 & PG0300 & 5.375 &  Ar  & 1.5 & 160 -- 530 &  500$\times$4 \cr
                   & 2021-11-05 &        &        &      & & & 600$\times$4 \cr
PKS 0837+012       & 2021-03-16 & PG0300 & 5.375 &  Ar  & 1.5 & 160 -- 530 &  500$\times$4  \cr
PKS 0907--023      & 2017-01-19 & PG0300 & 5.375 &  Ar  & 1.5 & 160 -- 530 &  600$\times$4 \cr
PKS 1034--293      & 2020-05-15 - & PG0300 & 5.375 &  Ar  & 1.5 & 160 -- 530 &  600$\times$4 \cr
                   & 2021-06-11   &        &        &      &              \cr
3C 273             & 2017-06-13 & PG0300 & 5.375 &  Ar  & 1.5 & 160 -- 530 &  20$\times$8  \cr
                   & 2017-06-14 &        &       &      &              \cr
PKS 1424--418      & 2022-07-26 & PG0900 & 12.875 & ThAr & 1.25 & 800 -- 1200 &  600$\times$4 \cr
                   & 2022-08-15 &        &        &      &              \cr
PKS 1510--089      & 2021-04-06 - & PG0900 & 12.875 & ThAr & 1.25 & 800 -- 1200 &  300$\times$4 \cr
                   & 2022-07-31   &        &        &      &              \cr
PKS 2023--07       & 2016-04-16 & PG0300 & 5.375 &  Ar  & 1.5 & 160 -- 530 &  300$\times$8 \cr
                   & 2018-10-04 &        &        &      &              \cr

\hline
\end{tabular}
\end{center}
\end{table*}

Data reduction was performed using a modified version of the pySALT/polSALT pipeline \citep{2010SPIE.7737E..25C,2022heas.confE..56C} where the wavelength calibration was performed using \textsc{iraf/noao}\footnote{\url{https://iraf.net/}}$^{,}$\footnote{\url{http://ast.noao.edu/data/software}} and enhanced cosmic ray rejection was implemented using the python-based \textsc{lacosmic} package.\protect\footnote{\url{https://github.com/larrybradley/lacosmic}} Due to the changing aperture size, absolute flux calibration of SALT spectroscopic data is not possible. However, relative flux calibration was performed for some of the sources with the use of spectrophotometric standard star observations. The standard star LTT 7987 was observed in longslit spectropolarimetry \textsc{linear} mode using the PG0300 grating at grating angles of $5.375^{\circ}$ and $5.75^{\circ}$, respectively, as well as the PG0900 grating at a grating angle of $12.875^{\circ}$.

\subsection{Optical photometry}
\label{subsec:specpol}

The optical photometric observations were taken with the Las Cumbres Observatory (LCO) Telescope Network's 1.0m telescopes, using the Sinistro CCD cameras at various observatories within the telescope network \citep{2013PASP..125.1031B}. These observations were taken as a quasi-contemporaneous complement to the SALT spectropolarimetric observations.

All pre-reductions were carried out using the automated \textsc{banzai}\footnote{\url{https://github.com/LCOGT/banzai}} pipeline. Then, standard aperture photometry was performed on the data with a python-based pipeline, using \textsc{astropy},\footnote{\url{https://www.astropy.org/}} \textsc{photutils},\footnote{\url{https://pypi.org/project/photutils/}} and \textsc{astroquery}.\footnote{\url{https://www.astropy.org/astroquery/}} The magnitudes were adjusted to the standard system by comparison to the AAVSO Photometric All Sky Survey (APASS) DR9 catalogue (accessed via Vizier\footnote{\url{https://vizier.u-strasbg.fr/viz-bin/VizieR-2}}). The resulting apparent magnitudes for each of the sources were then converted to flux,  using filter specific flux-zeropoint values \citep{2015PASP..127..102M, 1991PASP..103..661R}, as well as correcting for Galactic extinction \citep{2011ApJ...737..103S}.\footnote{\url{https://irsa.ipac.caltech.edu/applications/DUST/}} The errors quoted for the LCO data are statistical.

\subsection{Gamma-ray observations}
\label{subsec:gamma}

The $\gamma$-ray observations were obtained from the \textit{Fermi} Large Area Telescope (LAT). The \textit{Fermi}-LAT covers an energy range from 
$\sim 100$~MeV to $>$300\,GeV. It is a pair-conversion telescope with a large field of view and an angular resolution of $\sim$\,0.6$^{\circ}$ at 1\,GeV \citep{2009ApJS..183...46A,2009ApJ...697.1071A}. The publicly available data was accessed via the new \textit{Fermi} LAT Light Curve Repository (LCR) database \citep[][]{2023arXiv230101607A}. All of the $\gamma$-ray light curves presented in this paper were extracted from the LCR using 3-day binning, a minimum TS $\geq 4 (\gtrsim 2\sigma)$ for significant flux points, and a photon index fixed to the sources' 4FGL-DR2 catalog values \citep{2023ApJS..265...31A}. The errors quoted for the \textit{Fermi}-LAT data are the errors as given by the LCR database.

\section{Observational Results}
\label{sec:results}

In this section we highlight the observational results for five sources, (two BLLs and three FSRQs) that formed part of the larger sample of eighteen blazars. Three of the blazars, namely AP Lib, PKS 1034--293, and PKS 1510--089 \citep{2023ApJ...952L..38A}, were part of a long-term monitoring campaign. The remaining two blazars, PKS 0537--441 and 4FGL J0231.2--4745, formed part of a transient monitoring campaign and were observed around periods of increased $\gamma$-ray activity. The full set of results for the entire sample is summarised in Table~\ref{tab:full_results}.

\subsection{BL Lac-type objects}
\label{subsec:BLLs}

\subsubsection{PKS 0537--441}
\label{subsubsec:PKS0537}

PKS 0537--441 has a redshift of $z = 0.894$ and an average apparent visual magnitude of $V = 20.0 \pm 1.0$ during the time of our observations. It is an LBL with a synchrotron peak frequency of $\nu_{\rm sy} = 6.24 \times 10^{12}$\,Hz, as listed in the Fourth Catalog of Active Galactic Nuclei detected by the LAT \citep[4LAC;][]{2020ApJ...892..105A,2020arXiv201008406L}.

\begin{figure}
    \centering
    \includegraphics[width=\columnwidth]{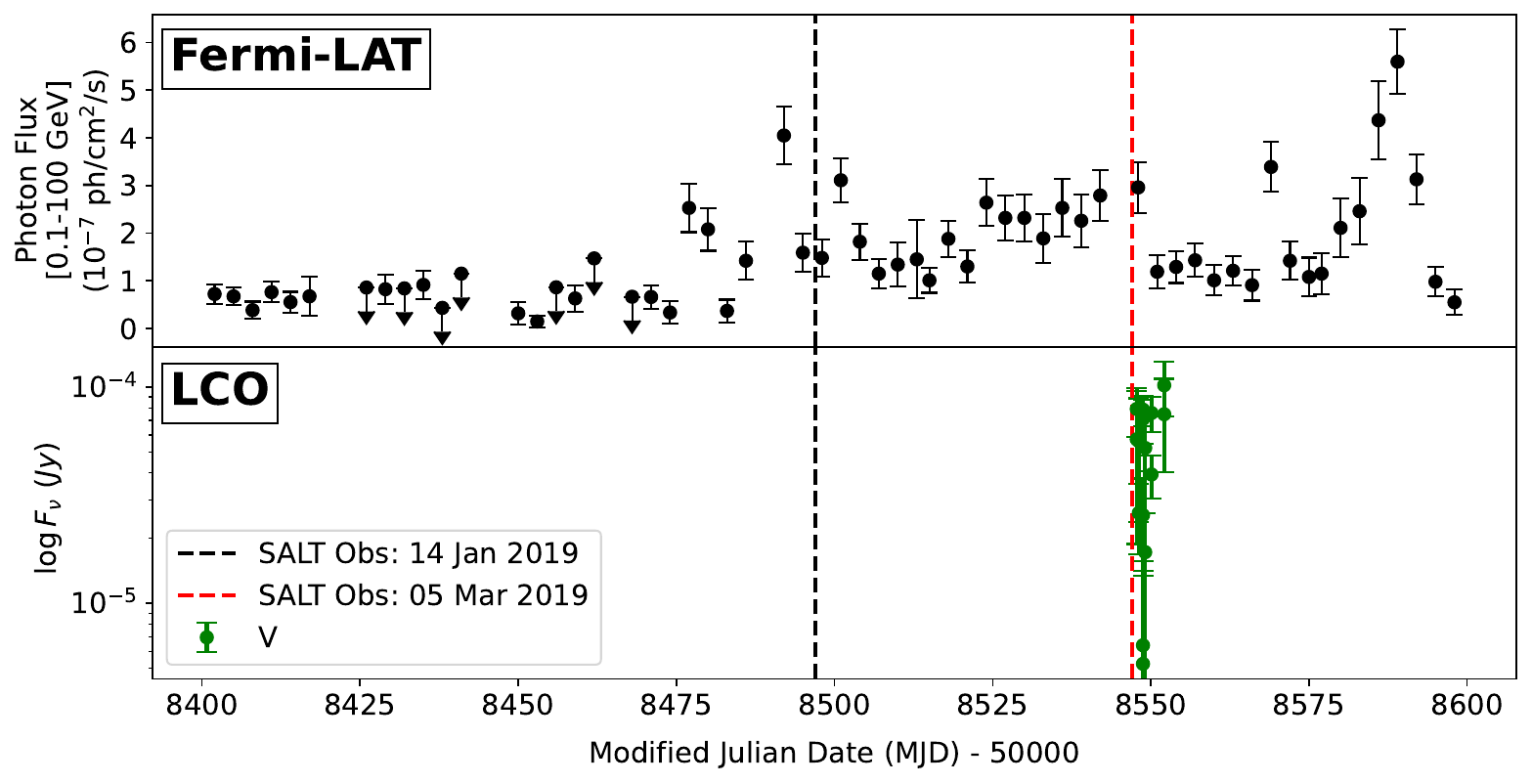}
    \caption{Gamma-ray (top panel) and optical (bottom panel) light curves for the BLL PKS 0537--441. The black and red dashed lines indicate the dates of the SALT spectropolarimetric observations.}
    \label{fig:PKS0537_lightcurves}
\end{figure}

In December 2018, PKS 0537--441 was increasing in brightness following an extended quiescent state \citep[e.g.][]{2019ATel12357....1N}. Fig.~\ref{fig:PKS0537_lightcurves} shows the $\gamma$-ray and optical light curves between 2018 October and 2019 April. The increase in activity can be seen in the $\gamma$-ray light curve around $\sim$\,MJD $58475$.

\begin{figure}
    \centering
    \includegraphics[width=\columnwidth]{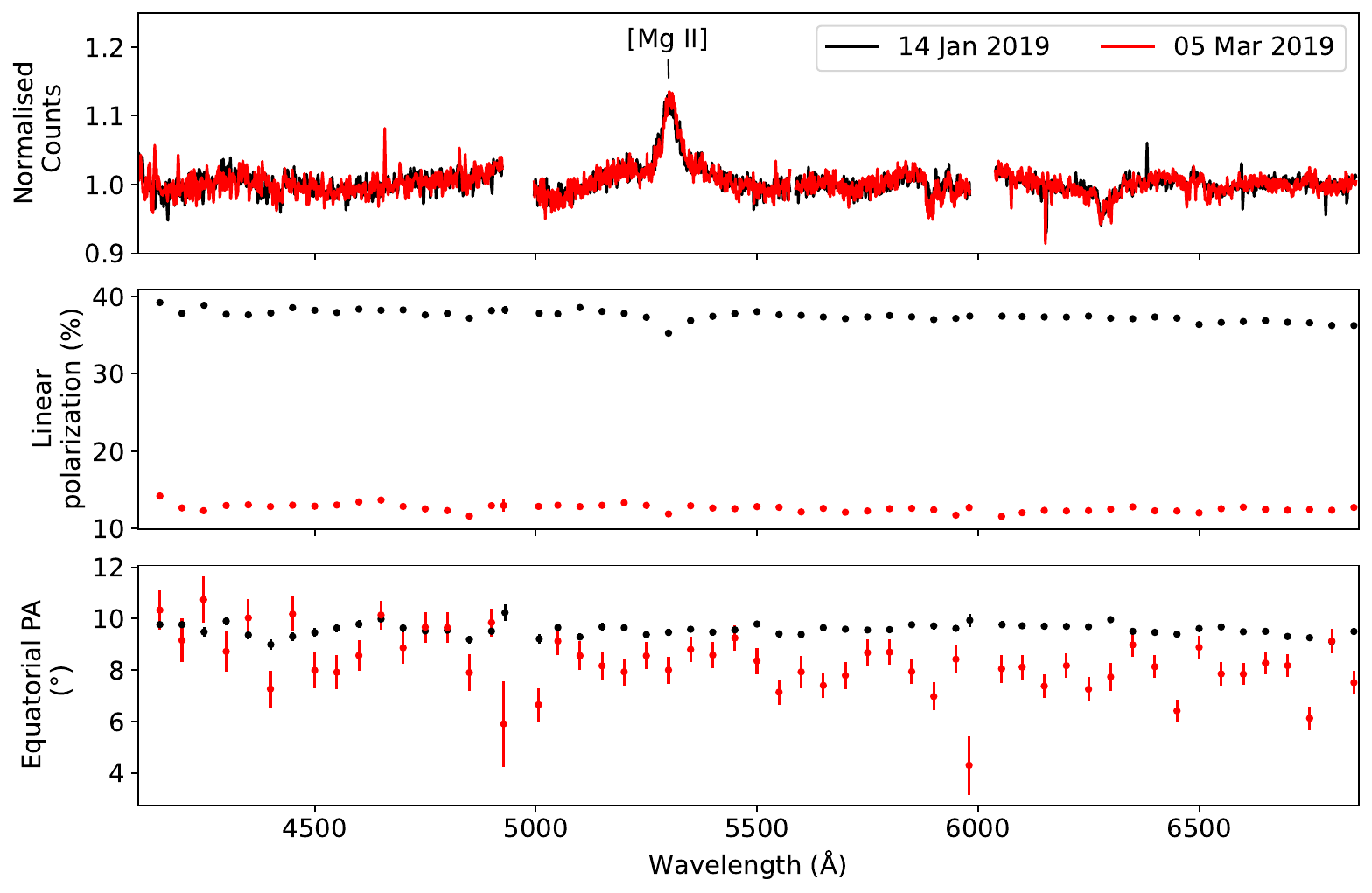}
    \caption{Spectropolarimetric observations of the BLL, PKS 0537–-441, observed on 2019 January 14 (black) and 2019 March 05 (red). The top panel shows the normalized counts spectra of the two observations, while the middle and bottom panels show the degree of linear polarization, and the equatorial polarization angle, respectively. The gaps in the spectra are due to the chip gaps of the CCD detector mosaic, and the removal of a strong skyline at $\lambda \approx 5580$\,\AA{}}
    \label{fig:PKS0537_specpol}
\end{figure}

Two SALT observations were taken on 2019 January 14, and on 2019 March 05. During the first observation, the source displayed a high average degree of polarization of $\langle\Pi\rangle = 37.5 \pm 0.7\,\%$ between $4100$\,{\rm \AA} and $6855$\,{\rm \AA} (Fig.~\ref{fig:PKS0537_specpol}). However, roughly three months later, the second observation showed a significantly lower degree of polarization of $\langle\Pi\rangle = 12.6 \pm 0.5\,\%$ ($4100$--$6855$\,{\rm \AA}). The equatorial polarization angle remained fairly constant between the two observations ($9.6 \pm 0.2^{\circ}$ and $8.4 \pm 1.0^{\circ}$, respectively), rotating by only $\sim 1^{\circ}$. The optical spectra display a forbidden [Mg II] emission line at $\lambda = 5303.5 \pm 1.5$\,{\rm \AA}. Despite the significant change in the degree of polarization, no significant change in the [Mg II] emission line strength was observed between the two observations, with the equivalent widths measured to be $|\rm W_\lambda| = 6.4 \pm 0.2$\,{\rm \AA} and $|\rm W_\lambda| = 6.5 \pm 0.2$\,{\rm \AA}, respectively. During the first observation, the isotropic-equivalent $\gamma$-ray luminosity was lower than during the second observation (with $\rm{L}_\gamma$ = ($3.9 \pm 1.1) \times 10^{47} \rm{erg s^{-1}}$ and $\rm{L}_\gamma$ = ($7.8 \pm 1.6) \times 10^{47} \rm{erg s^{-1}}$, respectively), indicating a ``disconnect'' between the $\gamma$-ray and optical behaviour during this active period. During both SALT observations, the degree of linear polarization exhibited a slight decrease towards longer wavelength (increasing towards higher frequencies).

\subsubsection{AP Lib}
\label{subsubsec:APLib}

\begin{figure}
    \centering
    \includegraphics[width=\columnwidth]{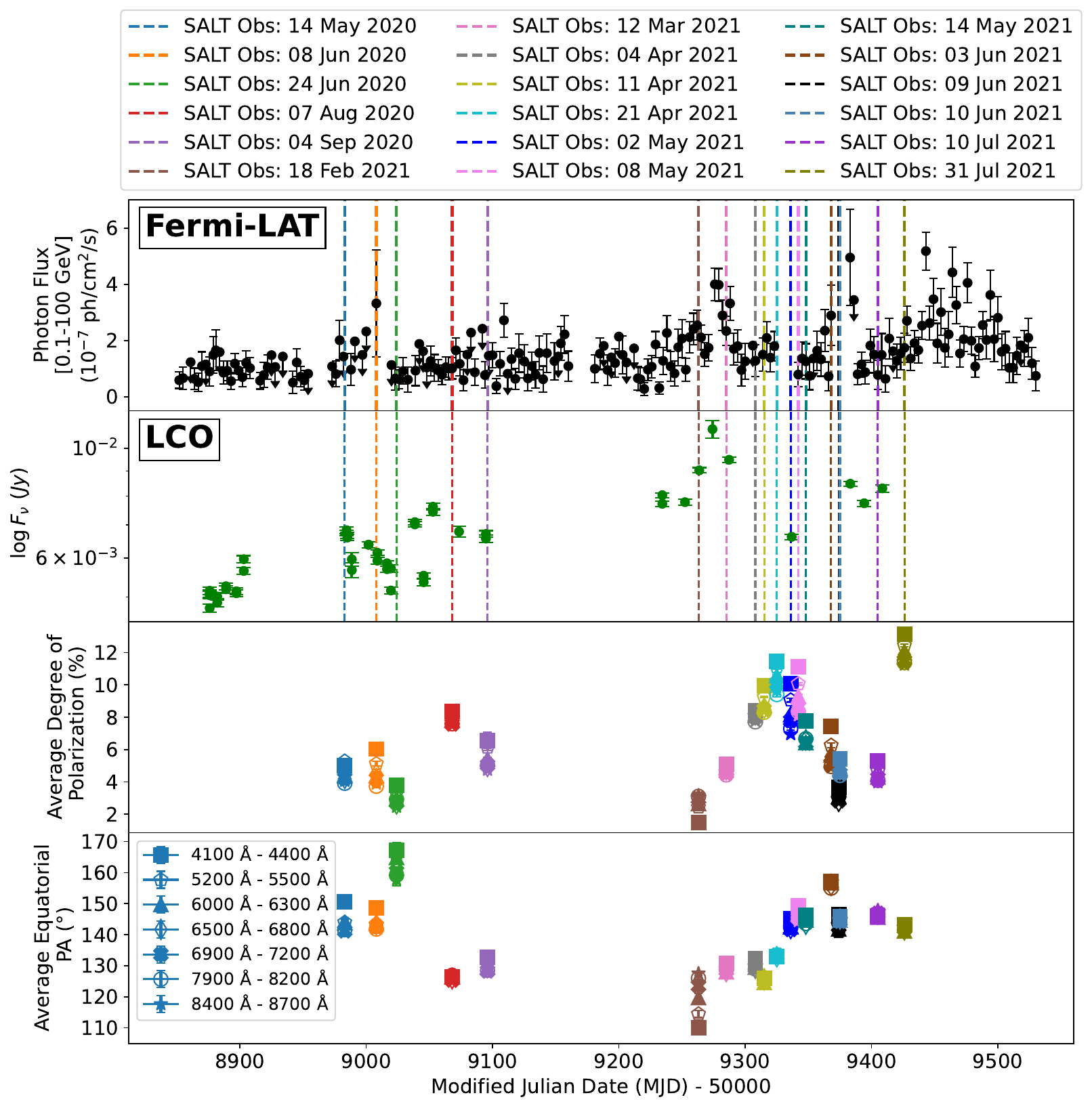}
    \caption{Gamma-ray (top panel) and optical (second panel) light curves of the BLL AP Lib, along with the averaged degree of polarization and equatorial polarization angle in seven different wavelength bands, reported for each SALT observation (third and bottom panel, respectively).}
    \label{fig:APLib_lightcurves}
\end{figure}

\begin{figure*}
    \centering
    \includegraphics[width=\textwidth]{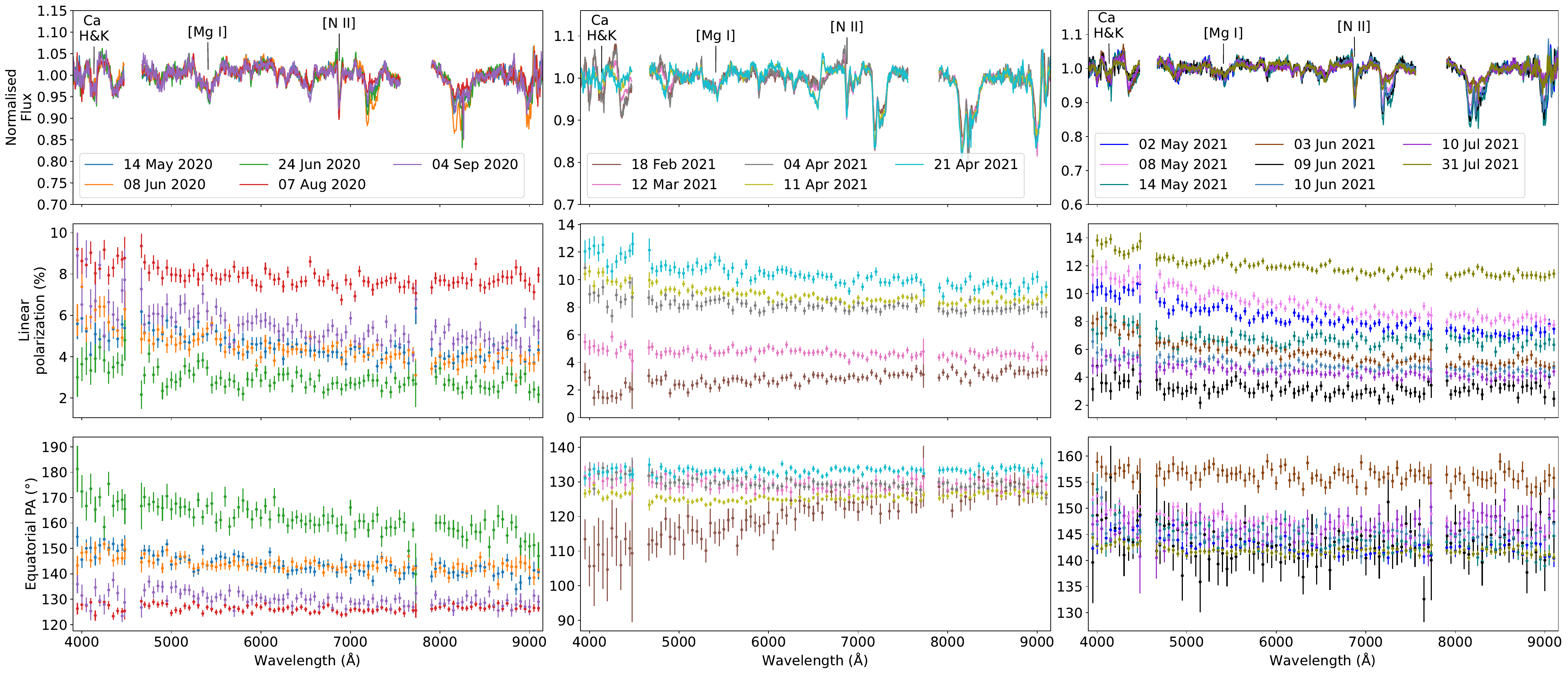}
    \caption{Spectropolarimetric observations of the BLL AP Lib, observed between 2020 May 14 and 2021 July 31, shown for 2020 (left column), 2021 Feb-April (middle column) and 2021 May - July (right column) for clarity. The top panels show the normalized flux spectra, while the middle and bottom panels show the degree of linear polarization, and the equatorial polarization angle for each observation, respectively. The gaps in the spectra are due to the chip gaps of the CCD detector mosaic, and the removal of a skyline at $\lambda = 7630$\,{\rm \AA}.}
    \label{fig:APLib_specpol}
\end{figure*}

AP Lib is a nearby BLL, with a redshift of $z = 0.049$, average apparent visual magnitude of $V = 14.8 \pm 0.2$ during the time of our observations, and a synchrotron peak frequency of $\nu_{\rm sy} = 9.06 \times 10^{13}$\,Hz. AP Lib was successfully observed eighteen times with SALT between 2020 May to 2021 July. The $\gamma$-ray and optical light curves during this period are shown in Fig.~\ref{fig:APLib_lightcurves}, along with the average degree of polarization and polarization angle in different wavelength ranges for each of the SALT observations. In both the $\gamma$-ray and optical light curves, the flux remained low up until $\sim$ MJD 59200, after which there is a marked increase in the flux which exhibited higher variability. During this period the degree of polarization fluctuated between $\langle\Pi\rangle = 12.0 \pm 0.7\,\%$ and $\langle\Pi\rangle = 2.7 \pm 0.6\,\%$ within the wavelength range of $\lambda = 3900 - 9150$\,{\rm \AA}. The equatorial polarization angle rotated by $42.5 \pm 0.7^{\circ}$. The degree of polarization appears to increase, with a time lag, after the increase in the $\gamma$-ray and optical fluxes around MJD 59500. A z-transformed discrete correlation function \citep[zDCF;][]{2014ascl.soft04002A} between the $\gamma$-ray light curve and the average degree of linear polarization showed that the increase in $\gamma$-ray flux occurred $54.1 \pm 2.6$ days before the observed increase in the degree of linear polarization. However, the spectropolarimetric observations are sparse just before this increase, and we cannot state with certainty that the time lag is real.

\begin{figure}
    \centering
    \includegraphics[width=\columnwidth]{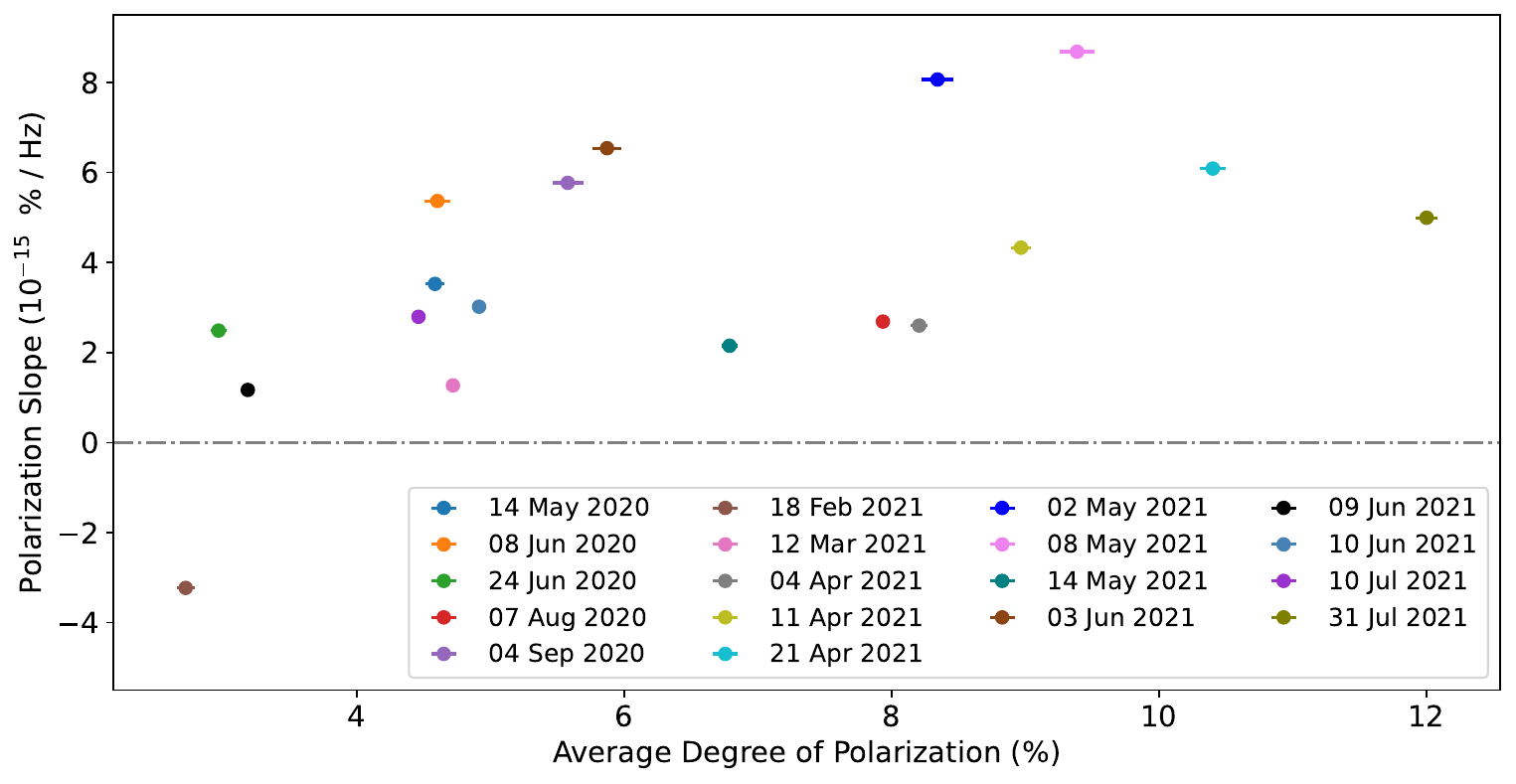}
    \caption{The frequency dependence of the degree of polarization as a function of the averaged degree of polarization for each of the observation dates, as observed for the BLL AP Lib. The dot-dashed grey line indicates where the slope of the frequency dependence is zero. A correlation of $\rho = 0.618$ (p-value $= 6.251 \times 10^{-3}$) is found (see text for details).}
    \label{fig:APLib_PolSlope_vs_Pol}
\end{figure}

The optical spectra and polarization are shown in Fig.~\ref{fig:APLib_specpol}. The degree of polarization increases towards higher frequencies for all but one observation. For the observation on 2021 February 18 (brown data points in the middle panel of Fig.~\ref{fig:APLib_specpol}), the polarization decreases towards higher frequencies (lower wavelengths). This negative trend in the frequency dependence of the polarization is seen in the date for which the lowest degree of polarization is observed.
For AP Lib, the frequency dependence of the polarization becomes increasingly positive as the degree of polarization increases (see Fig.~\ref{fig:APLib_PolSlope_vs_Pol}), with a Spearman's test showing a strong, significant correlation of $\rho = 0.618$ (p-value $= 6.251 \times 10^{-3}$).

The possible time-lag between the light curves (both optical and $\gamma$-ray) and the average degree of polarization, along with the frequency dependence that changed between 2021 February 18 and the next observation, may indicate that a large scale change in the jet occurred. However, since the observational data is limited, further investigation and statistical analyses will be required to confirm this hypothesis.

\subsection{Flat-spectrum radio quasars}
\label{subsec:FSRQs}

\subsubsection{4FGL J0231.2--4745}
\label{subsubsec:4FGLJ0231}

4FGL J0231.2--4745 is a faint FSRQ with an average apparent visual magnitude of $V = 17.0 \pm 0.3$ during the time of our observations, a redshift of $z = 0.765$, and $\nu_{\rm sy} = 1.31 \times 10^{12}$\,Hz. Enhanced $\gamma$-ray activity was detected for this source on 2019 October 17 \citep{2019ATel13209....1P}. It reached its highest ever average daily flux on 2019 October 19, increasing by a factor of $60$ when compared to the average flux from the 4FGL \textit{Fermi}-LAT catalogue \citep{2019ATel13209....1P}.

\begin{figure}
    \centering
    \includegraphics[width=\columnwidth]{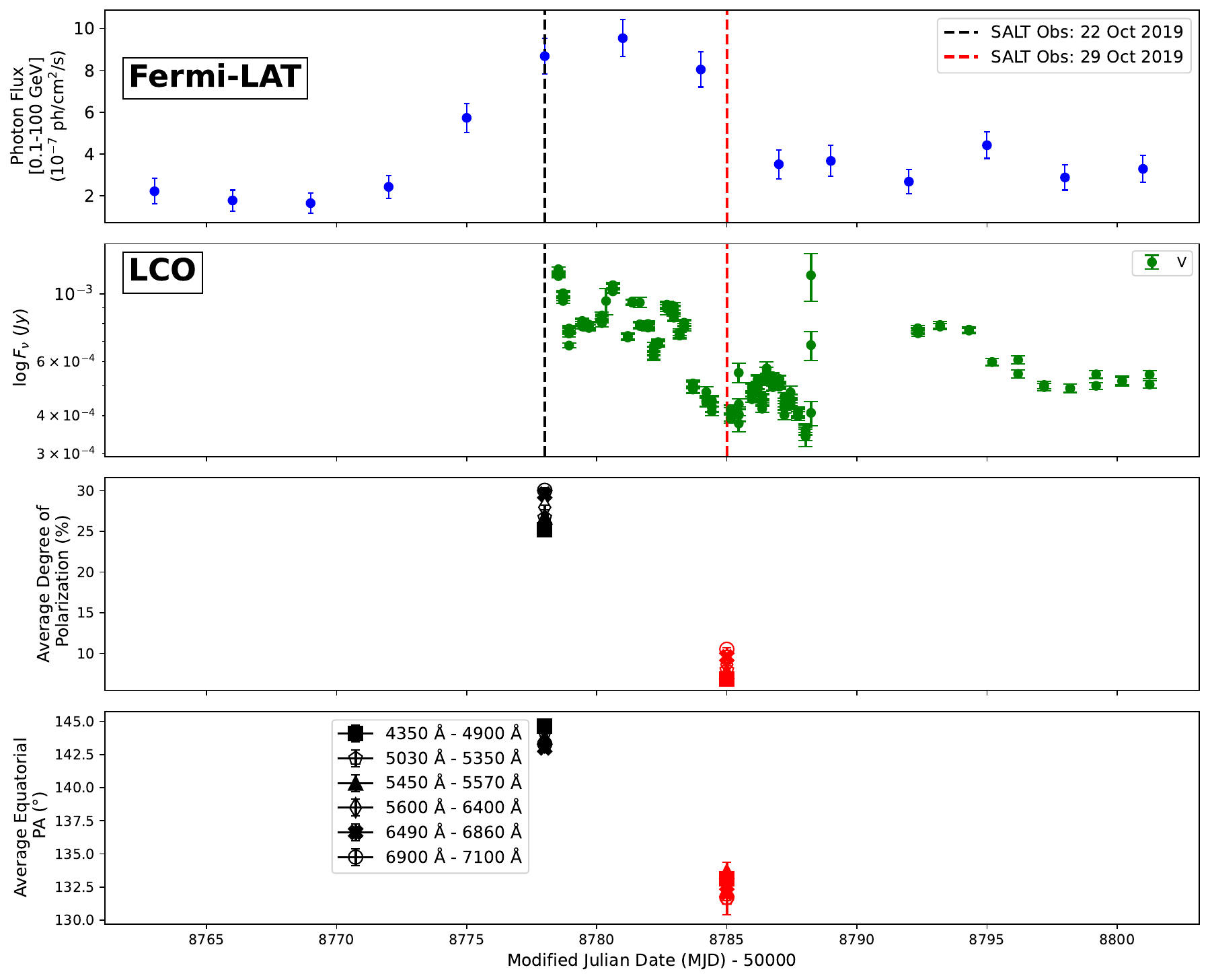}
    \caption{Gamma-ray (top panel) and optical (bottom panel) light curves of the FSRQ 4FGL J0231.2--4745. The black and red dashed lines indicate the dates of the SALT spectropolarimetric observations.
    }
    \label{fig:4FGLJ0231_lightcurves}
\end{figure}

\begin{figure}
    \centering
    \includegraphics[width=\columnwidth]{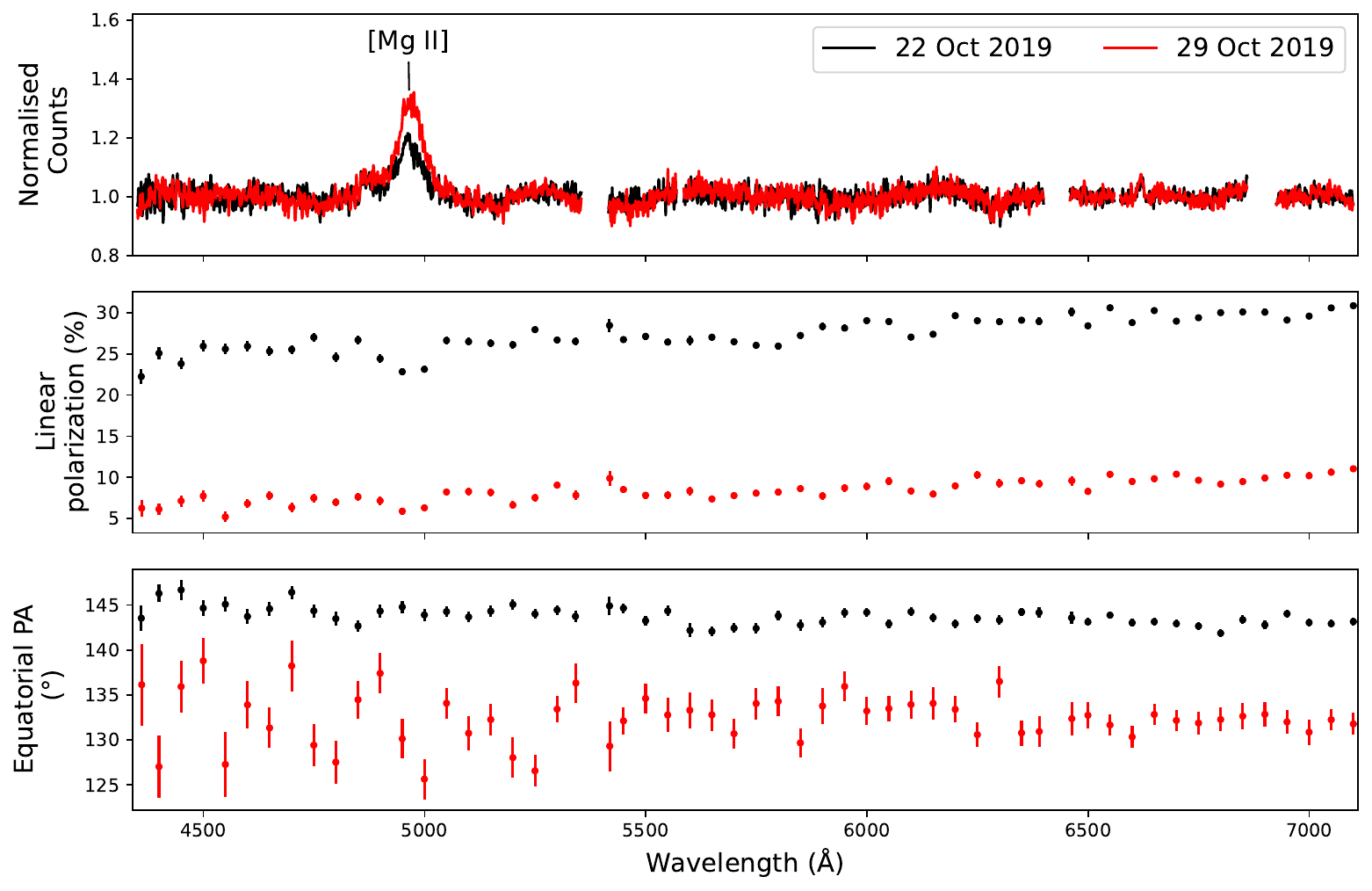}
    \caption{Spectropolarimetric observations of the FSRQ 4FGL J0231.2--4745 on 2019 October 22 (black) and 2019 October 29 (red). The top panel shows the normalized counts spectra of the two observations, while the middle and bottom panels show the degree of linear polarization and the equatorial polarization angle, respectively. The gaps in the spectra are due to the chip gaps of the CCD detector mosaic, and the removal of two strong skylines at $\lambda = 5580$\,{\rm \AA} and $\lambda = 6566$\,{\rm \AA}, respectively.}
    \label{fig:4FGLJ0231_specpol}
\end{figure}

The $\gamma$-ray and optical light curves during the 2019 October flare are shown in Fig.~\ref{fig:4FGLJ0231_lightcurves}. The $\gamma$-ray flare lasted from $\sim$\,MJD $58770$ to MJD $58790$. The spectropolarimetric observations were taken on 2019 October 22 and 2019 October 29, one before the peak of the flare, and one after (see Fig.~\ref{fig:4FGLJ0231_lightcurves}), and the results are shown in Fig.~\ref{fig:4FGLJ0231_specpol}. During the first observation (prior to the peak, as the flare was rising) the average degree of polarization was $\langle\Pi\rangle = 27.5 \pm 1.9\,\%$ ($\lambda = 4350 - 7100$\,{\rm \AA}); during the second, post-peak observation, the source exhibited a decreased average degree of polarization of $\langle\Pi\rangle = 8.4 \pm 1.3\,\%$. The polarization angle changed from $143.8 \pm 1.0^{\circ}$ to $132.6 \pm 2.8^{\circ}$ between the two observations.

The optical spectra exhibit a [Mg II] emission line at $\lambda = 4966.1 \pm 6.7$\,{\rm \AA}, with equivalent widths increasing from $|W_\lambda| = 11.8 \pm 1.0$\,{\rm \AA} on October 22, to $|W_\lambda| = 21.3 \pm 1.2$\,{\rm \AA} on October 29, indicating that the non-thermal emission contribution had diminished greatly between the two spectropolarimetric observations. For both of the SALT observations, the degree of polarization decreased towards higher frequencies.

\subsubsection{PKS 1034--293}
\label{subsubsec:PKS1034}

\begin{figure}
    \centering
    \includegraphics[width=\columnwidth]{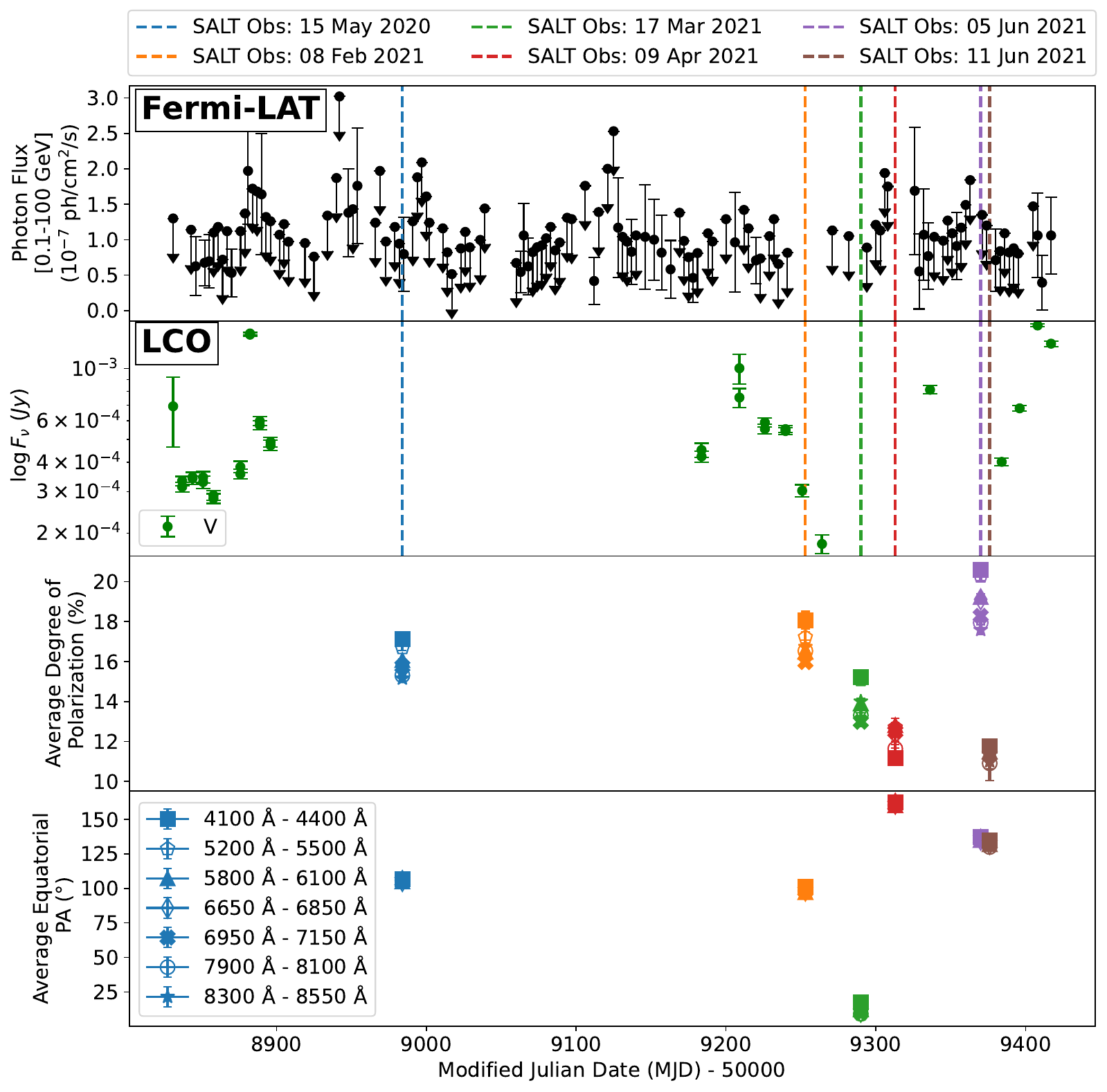}
    \caption{Gamma-ray (top panel) and optical (second panel) light curves of the FSRQ PKS 1034--293, along with the average degree of polarization and equatorial polarization angle in seven different wavelength bands, reported for each SALT observation (third and bottom panel, respectively).}
    \label{fig:PKS1034_lightcurves}
\end{figure}

PKS 1034--293 is an FSRQ with a redshift of $z = 0.310$, an average apparent visual magnitude of $V = 17.3 \pm 0.6$ during the time of our observations, and synchrotron peak frequency of $\nu_{\rm sy} = 6.92 \times 10^{12}$\,Hz. It has been successfully observed six times with SALT between 2020 May and 2021 July as part of a long-term monitoring campaign. The $\gamma$-ray and optical light curves for this period are shown in Fig.~\ref{fig:PKS1034_lightcurves}, along with the average degree of polarization and polarization angle for each of the SALT observations in different wavelength ranges. The $\gamma$-ray and optical fluxes remained fairly low throughout the reported period, and the majority of the LAT data points were upper limits only.

\begin{figure}
    \centering
    \includegraphics[width=\columnwidth]{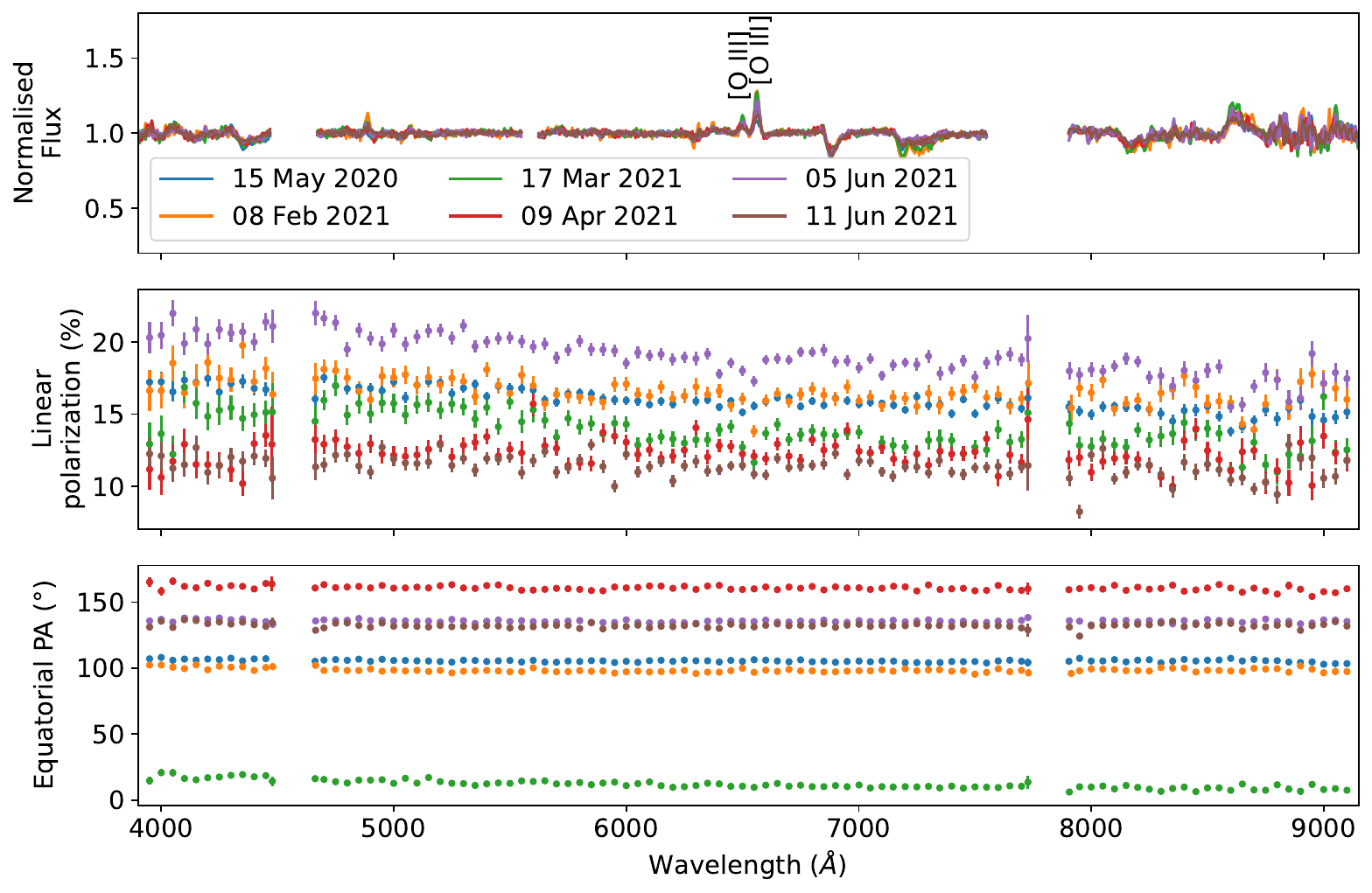}
    \caption{Spectropolarimetric observations of the FSRQ PKS 1034--293 between 2020 May 15 and 2021 June 11. The top panel shows the normalized counts spectra of the two observations, while the middle and bottom panels show the degree of linear polarization, and the equatorial polarization angle, respectively. The gaps in the spectra are due to the chip gaps of the CCD detector mosaic, and the removal of two strong skylines at $\lambda = 5580$\,{\rm \AA} and $\lambda = 7625$\,{\rm \AA}, respectively.}
    \label{fig:PKS1034_specpol}
\end{figure}

The average degree of polarization reached a maximum of $\langle\Pi\rangle = 19.1 \pm 1.3\,\%$, and a minimum of $\langle\Pi\rangle = 11.5 \pm 0.1\,\%$ (in the wavelength range of $\lambda = 3900 - 9150$\,{\rm \AA}). The equatorial polarization angle exhibited high variability, rotating through a range of $\sim\,150^{\circ}$ (from $160.7 \pm 1.9^{\circ}$ to $12.0 \pm 3.1^{\circ}$). Two forbidden [O III] lines were observed in the optical spectra (Fig.~\ref{fig:PKS1034_specpol}) at $\lambda = 6500.0 \pm 9.2$\,{\rm \AA} and $\lambda = 6561.9 \pm 5.1$\,{\rm \AA}, and remained fairly constant throughout the observing period, as can be expected due to the unchanging non-thermal emission.

\begin{figure}
    \centering
    \includegraphics[width=\columnwidth]{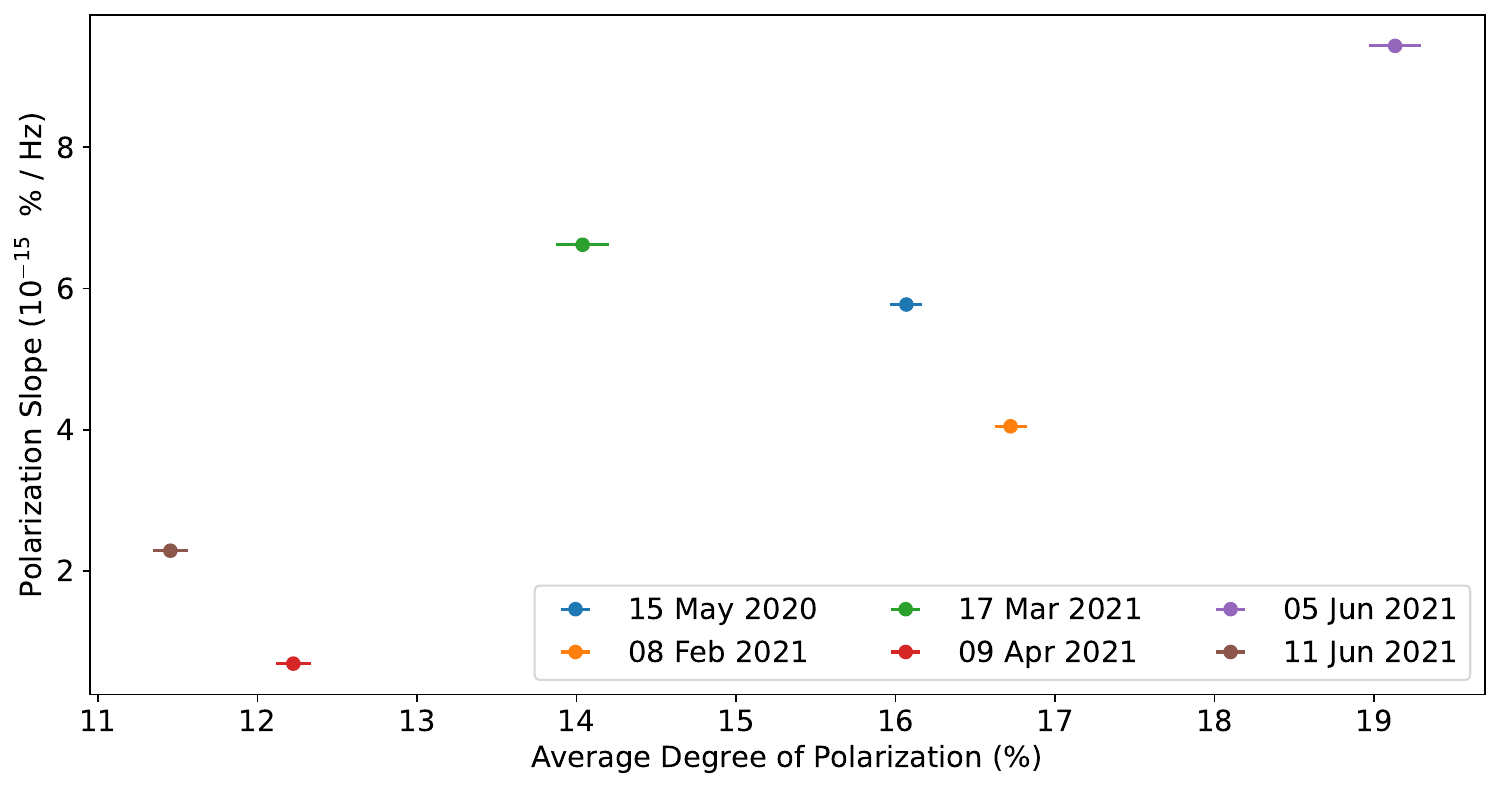}
    \caption{The frequency dependence of the degree of polarization as a function of the average degree of polarization for each of the observation dates, as observed for the FSRQ PKS 1034--293. A Spearman's test gives, $\rho = 0.714$ (p-value = 0.111), see text for details}.
    \label{fig:PKS1034_PolSlope_vs_Pol}
\end{figure}

During all of the SALT observations, the degree of polarization increased towards higher frequencies (Fig.~\ref{fig:PKS1034_PolSlope_vs_Pol}). While, by eye, there seems to be a trend where the wavelength dependence of the degree of polarization increases with the degree of polarization (similar to what was found for AP~Lib; Fig.~\ref{fig:APLib_PolSlope_vs_Pol}), this is not statistically significant (Spearman's test gives $\rho = 0.714$, with p-value = 0.111).

\subsubsection{PKS 1510--089}
\label{subsubsec:PKS1510}

\begin{figure}
    \centering
    \includegraphics[width=\columnwidth]{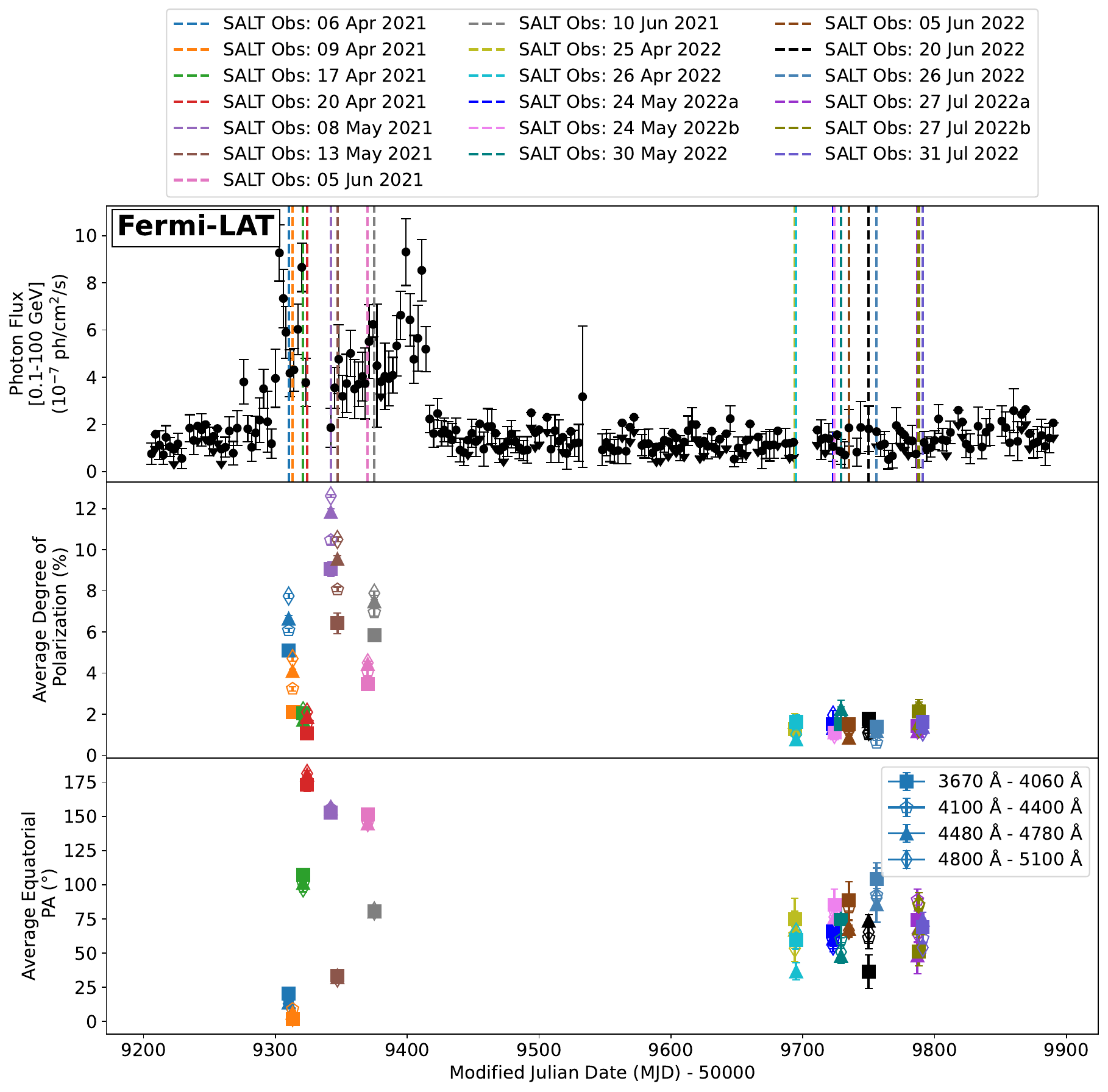}
    \caption{Gamma-ray (top panel) light curve of the FSRQ PKS 1510--089, along with the average degree of polarization and equatorial polarization angle in four different wavelength bands, reported for each SALT observation (middle and bottom panel, respectively).}
    \label{fig:PKS1510_lightcurves}
\end{figure}

PKS 1510--089 is situated at a redshift of $z = 0.36$ with a low-state apparent visual magnitude of $V = 16.54$ and synchrotron peak frequency of $\nu_{\rm sy} = 1.10\times10^{13}$\,Hz. It displayed highly variable emission in the $\gamma$-ray regime during the first eight years of \textit{Fermi}-LAT monitoring, but has since been less active. Even during low states, this source emits strong VHE (very high energy) emission. The persistent VHE emission from PKS 1510--089 could possibly be explained by hadronic processes \citep{2018heas.confE..33Z, 2022MNRAS.515.5242D} or by the inclusion of a second emission zone far from the central engine, dominating the persistent X-ray and VHE $\gamma$-ray emission \citep{2023ApJ...952L..38A}. The \textit{Fermi}-LAT $\gamma$-ray flux suddenly dropped in July 2021 and remained in a near constant low state since then (Fig.~\ref{fig:PKS1510_lightcurves}). In addition to the drop in the high-energy $\gamma$-ray flux, the optical flux and polarization also vanished \citep{2023ApJ...952L..38A}. Part of the optical results presented here were originally included in \citet{2023ApJ...952L..38A}, but this paper presents expanded optical spectropolarimetry results.

\begin{figure*}
    \centering
    \includegraphics[width=\textwidth]{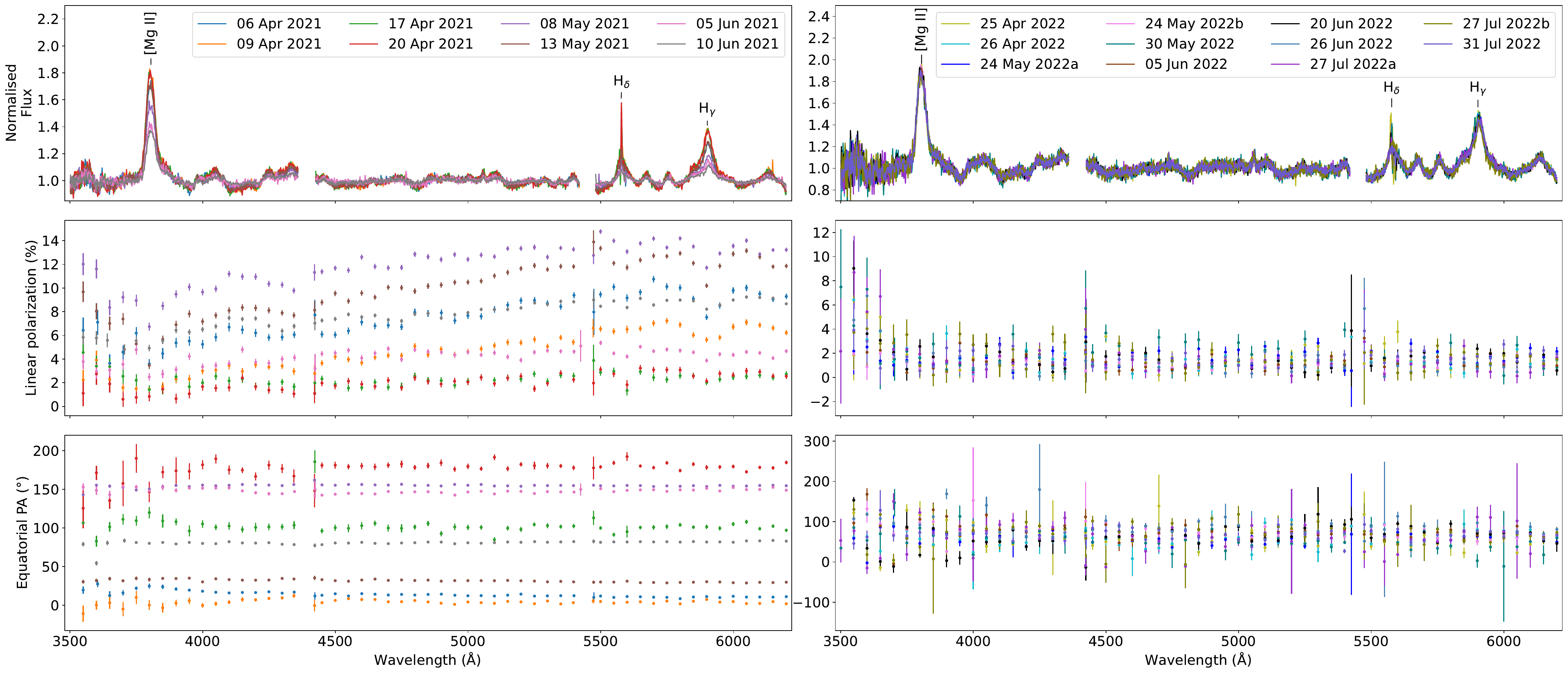}
    \caption{Spectropolarimetric observations of the FSRQ PKS 1510--089 between 2021 April 06 and 2022 July 31, shown separately for 2021 (left column) and 2022 (right column). The top panels show the normalized counts spectra, where the middle and bottom panels show the degree of linear polarization and the equatorial polarization angle, respectively. The gaps in the spectra are due to the chip gaps of the CCD detector mosaic.}
    \label{fig:PKS1510_specpol}
\end{figure*}

PKS 1510--089 was observed successfully with SALT nineteen times between 2021 April to 2022 July. Of these, eight observations were taken during 2021, and eleven during 2022. During the 2021 observing season, the source exhibited significant flaring activity in the $\gamma$-ray regime. The average degree of polarization reached a maximum of $\langle\Pi\rangle = 12.5 \pm 1.1\,\%$ and a minimum of $\langle\Pi\rangle = 2.2 \pm 0.5\,\%$ (between $\lambda = 4100 - 6200$\,{\rm \AA}). The polarization angle varied substantially, ranging from $4.7 \pm 2.7^{\circ}$ to $178.9 \pm 4.8^{\circ}$ (see the left panels of Fig.~\ref{fig:PKS1510_specpol}).

During the 2022 observing season, the source showed little to no activity in the $\gamma$-ray regime. This is mirrored by the behaviour in the optical regime (see the right panels of Fig.~\ref{fig:PKS1510_specpol}). During this period, the strength and width of the emission features remained unchanged, and the degree of linear polarization remained below $2\,\%$ for the entire observing period. The polarization angle remained more constant during this period than during 2021, ranging from $83.9 \pm 22.9^{\circ}$ to $49.0 \pm 13.9^{\circ}$. The polarization levels during 2022 are comparable to that of a comparison star on the slit (with $\langle \Pi \rangle_{\rm comp} = 1.3 \pm 0.7\,\%$), and is therefore likely not intrinsic to the source itself.

The optical spectra show three emission lines: [Mg II] at $\lambda = 3804.4 \pm 1.5$\,{\rm \AA}, H$_{\delta}$ at $\lambda = 5578.8 \pm 1.6$\,{\rm \AA}, and H$_\gamma$ at $\lambda = 5904.5 \pm 0.9$\,{\rm \AA}. The equivalent widths of the emission lines during the 2022 season were, on average, much larger than in 2021, indicating the diminished non-thermal contribution in 2022 \citep{2023ApJ...952L..38A}.

For the 2021 observations, the degree of linear polarization was observed to decrease towards higher frequencies. As no significant intrinsic polarization was measured in 2022, no statements about its frequency dependence in that observing season can be made.

\begin{figure}
    \centering
    \includegraphics[width=\columnwidth]{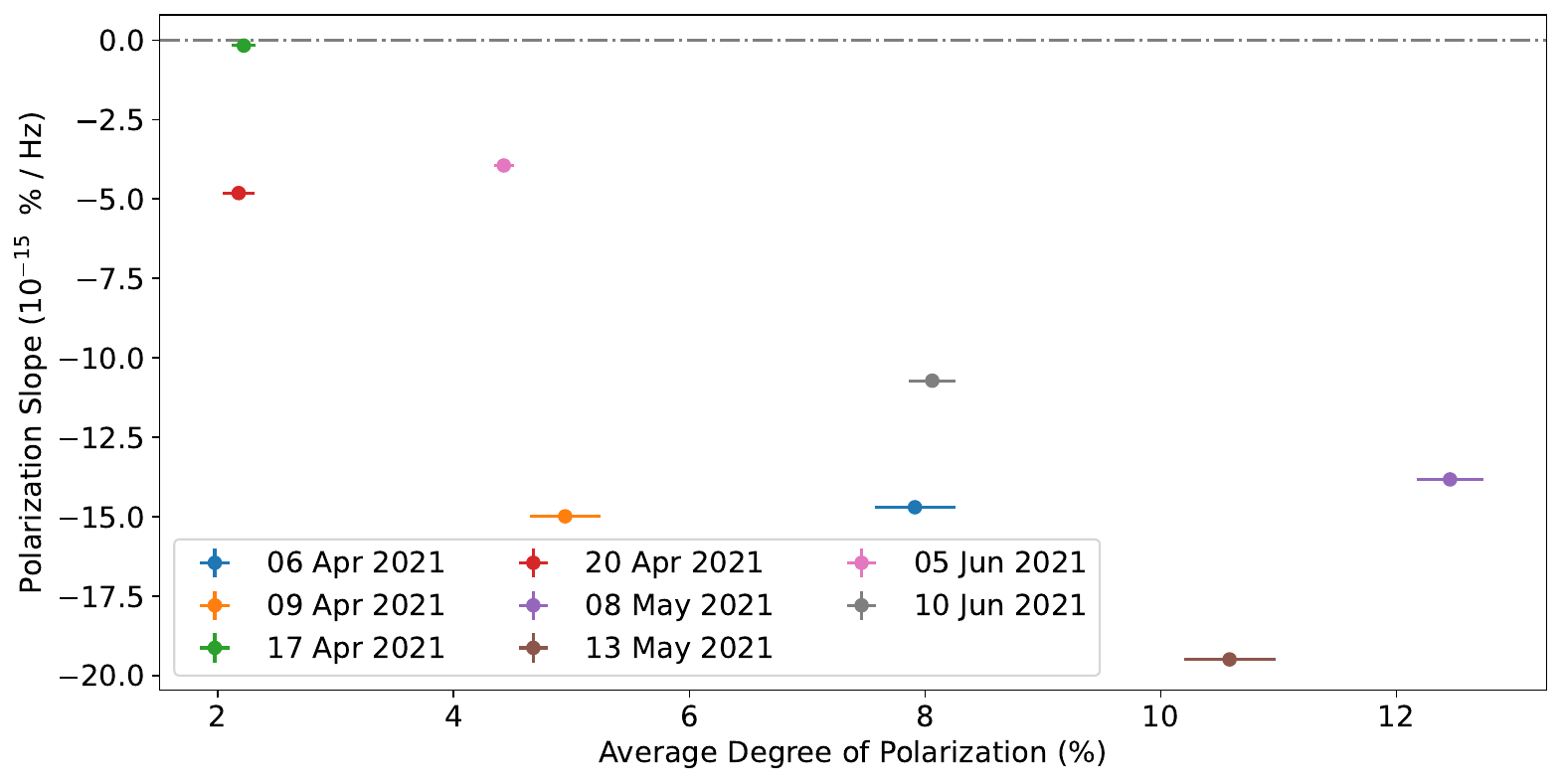}
    \caption{The frequency dependence of the degree of polarization as a function of the average degree of polarization for each of the observation dates, as observed for the FSRQ PKS 1510--089 during the 2021 semester. The dotted-dashed grey line indicates where the frequency dependent slope is zero. A correlation of $\rho = -0.782$ (p-value = $7.519 \times 10^{-5}$) is found if all points are considered; see text for details.}
    \label{fig:PKS1510_PolSlope_vs_Pol}
\end{figure}

For the data from 2021, there is a general trend for the frequency dependence of the degree of polarization to become increasingly negative with increasing degree of polarization (Fig.~\ref{fig:PKS1510_PolSlope_vs_Pol}). A Spearman's test on the 2021 data resulted in strong and significant anti-correlation of $\rho = -0.782$ (p-value = $7.519 \times 10^{-5}$). However, we caution that the data points for 2021 April 17 and 20 may affect the credibility of correlation, as the average polarization of PKS 1510--089 on those days were comparable to that of the comparison star ($\langle \Pi \rangle = 2.22\,\%$ and $2.18\,\%$, respectively). If these two data points are not included, the anti-correlation found is weaker and far less significant: $\rho = -0.314$ (p-value = 0.544).
Thus, the correlation found for PKS 1510--089 during 2021 should be treated with caution. If this trend is intrinsic to the source itself, it is the opposite of what has been found for AP Lib (Section \ref{subsubsec:APLib}) and PKS 1034--293 (Section \ref{subsubsec:PKS1034}).

The change in the frequency dependence of the degree of polarization between 2021 and 2022, the sudden decrease of the high-energy activity (Fig.~\ref{fig:PKS1510_lightcurves}), and the low polarization levels during the 2022 period, indicate that the source is in an unprecedented low-state, with little (if any) non-thermal contribution to the observed optical emission. The historically active HE $\gamma$-ray emitting jet-component seems to have been completely diminished. \citet{2023ApJ...952L..38A} suggested that there were two separate emission regions actively contributing to the observed multi-wavelength flux and that the primary emission zone responsible for the high-energy $\gamma$-ray emission vanished, leaving behind only the secondary zone that dominates the VHE $\gamma$-ray and X-ray emission, because either the inner jet has weakened, or is has swung away from the line of sight.

\section{Discussion}
\label{sec:discussion}

In total, six BLLs and twelve FSRQs were observed. Of the six BLLs, three were LBLs ($\nu_{\rm sy} < 10^{14}$\,Hz), two were IBLs ($10^{14} < \nu_{\rm sy} < 10^{15}$\,Hz), and one was an HBL ($\nu_{\rm sy} > 10^{15}$\,Hz). All twelve FSRQs were LSPs, with $\nu_{\rm sy} < 10^{14}$\,Hz. Of the eighteen blazars investigated, three were observed only during low states, eleven were observed only during high states, and four were observed during both high and low states. This section will discuss statistical trends found for these blazars from the observational results presented in Section \ref{sec:results} and Appendix \ref{app:full_results}.

\begin{figure}
    \centering
    \includegraphics[width=\columnwidth]{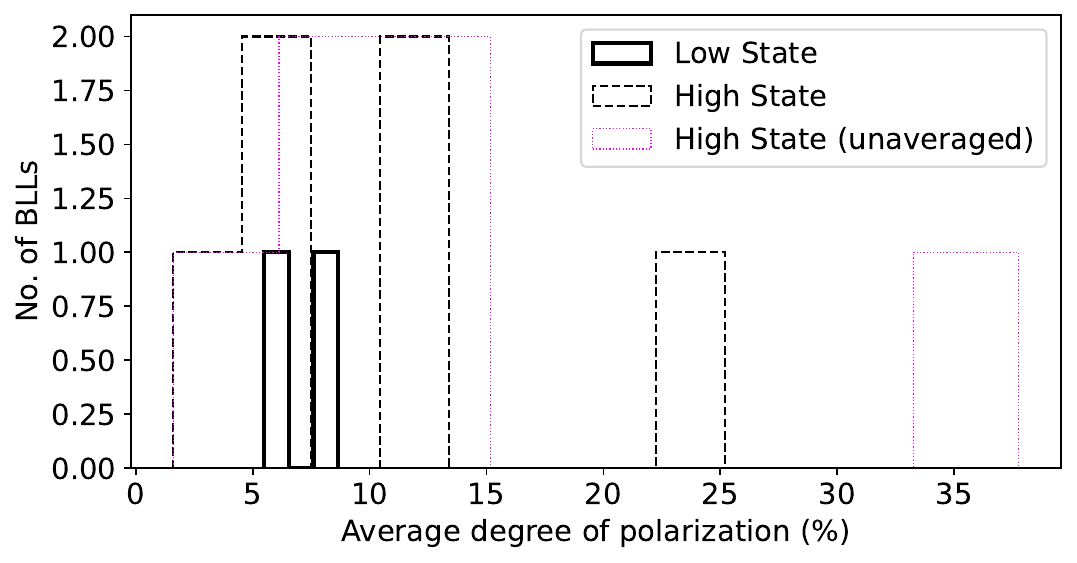}
    \caption{The average degree of polarization of the entire BLL sample. The polarization levels for low/quiescent states are indicated by the solid black line, whereas the high/active states are indicated by the dashed black line. The dashed magenta line indicates the distribution when only the maximum degree of polarization is considered for PKS 0537--441.}
    \label{fig:Pol_BLL}
\end{figure}

\begin{figure}
    \centering
    \includegraphics[width=\columnwidth]{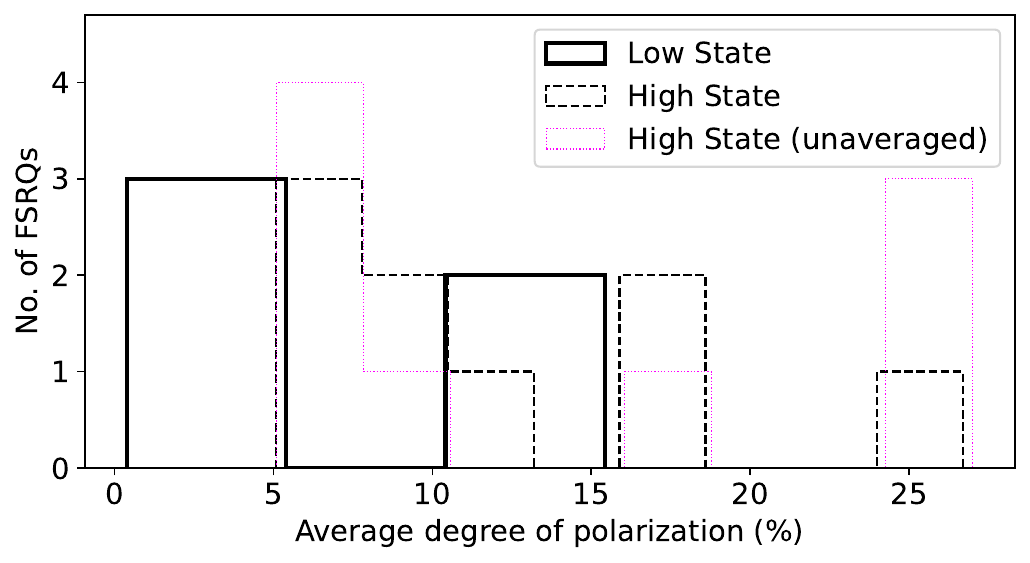}
    \caption{The average degree of polarization of the entire FSRQ sample. The polarization levels for low/quiescent states are indicated by the solid black line, whereas the high/active states are indicated by the dashed black line. The dashed magenta line indicates the distribution when only the maximum degree of polarization is considered for PKS 0208--512 and 4FGL J0231.2--4745.}
    \label{fig:Pol_FSRQ}
\end{figure}

The distribution of the average degree of polarization (between $\lambda = 4100 - 6200$\,{\rm \AA}) for the sample of BLLs and FSRQs for low and high states are shown in Figs.~\ref{fig:Pol_BLL} and \ref{fig:Pol_FSRQ}, respectively. For three sources (PKS 0537--441, PKS 0208--512, and 4FGL J0231.2--4745), the degree of polarization changed by more than $15\%$ during the same high state. The dashed magenta lines in Figs.~\ref{fig:Pol_BLL} \& \ref{fig:Pol_FSRQ} show the distribution if only the higher value is considered for these sources. 

The spread/standard deviation in the degree of polarization during low states over all of the sources was $\sigma \approx 5$~\%. During high states, the degree of linear polarization had a significantly larger spread in the data, with a standard deviation of $\sigma \approx 7$~\%. For the BLLs, the average degree of polarization during a low state of activity reached a maximum of $\Pi = 8.7 \pm 0.4\,\%$. For the high states the maximum average degree of polarization reached was $\Pi = 25.0 \pm 0.5\,\%$, but for most sources remained below $15\%$. For the FSRQs, the average degree of polarization reached a maximum of $\Pi = 15.4 \pm 2.8\%$ during low states, and $\Pi = 26.7 \pm 0.4\%$ during high states. While on average, the degree of polarization during high states is higher for FSRQs than for BLLs, more observations were taken of FSRQs in $\gamma$-ray active states than during quiescence (due to the nature of the transient observing campaign). A Levene test and t-test showed no statistically significant difference between the two samples.

For the majority of the sources observed multiple times in $\gamma$-ray high states, the average degree of polarization, $\langle \Pi \rangle$, was higher while the $\gamma$-ray fluxes were still on the rise, and lower while the $\gamma$-ray fluxes were falling, even if the level of flux was similar. This was seen especially for three sources: PKS 0537--441, PKS 0208--512, and 4FGL J0231.2--4745. To a lesser extent, it could also be seen for PKS 1424--418. Even though this trend was not observed for all of the sources, it might point to an underlying physical trend/similarity, and warrants further investigation in future work.

\subsection{Statistical trends in the data}
\label{subsec:Trends}

\begin{figure}
    \centering
    \includegraphics[width=\columnwidth]{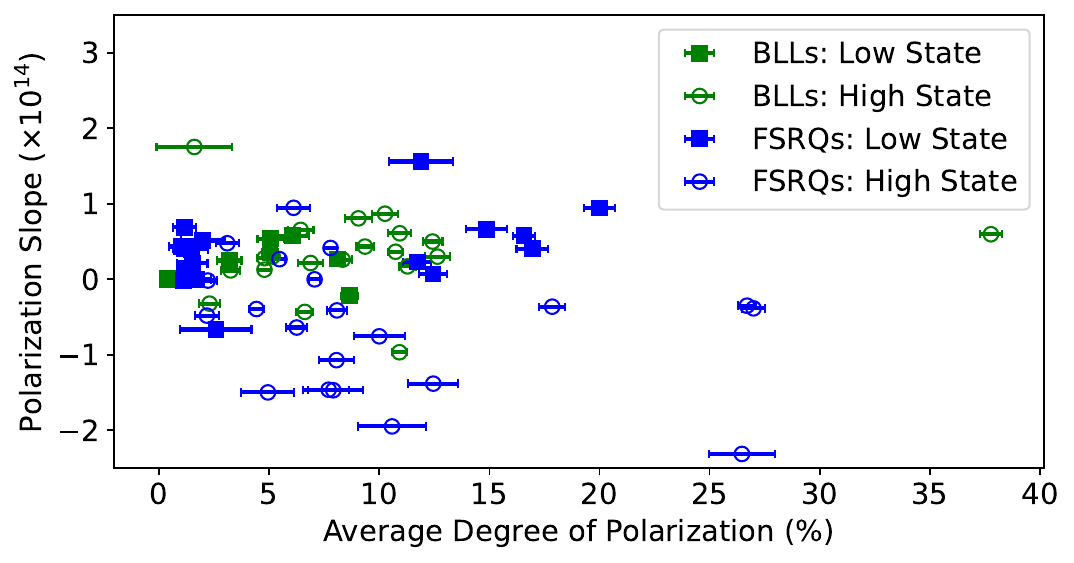}
    \caption{The relation between the degree of polarization and its frequency dependence for all of the sources in the sample. The open circles indicate the averaged degree of linear polarization of blazars for high/active states, and the filled squares indicate the averaged degree of linear polarization of blazars for low/quiescent states. Green data points indicate values associated with BLLs, and blue data points indicate values associated with FSRQs.}
    \label{fig:PolSlope_vs_Pol}
\end{figure}

As discussed in Section\,3, trends in the frequency dependence of the polarization as it changes with the average degree of polarization were found for the three sources, 
namely AP Lib (Section \ref{subsubsec:APLib}), PKS1034--293 (Section \ref{subsubsec:PKS1034}), and PKS 1510--089 (Section \ref{subsubsec:PKS1510}). As these showed some correlation, the idea was expanded to include all of the data from all of the sources (see Fig.~\ref{fig:PolSlope_vs_Pol}). No discernible trend in the frequency dependence of the degree of polarization and the average degree of polarization was found for the entire sample for high states, with $\rho = -0.080$ (p-value = 0.621). For low states, a slightly more significant correlation of $\rho = 0.375$ (p-value = 0.054) was found, i.e., a slight trend for the polarized emission becoming redder with increasing degree of polarization.

\begin{figure}
    \centering
    \includegraphics[width=\columnwidth]{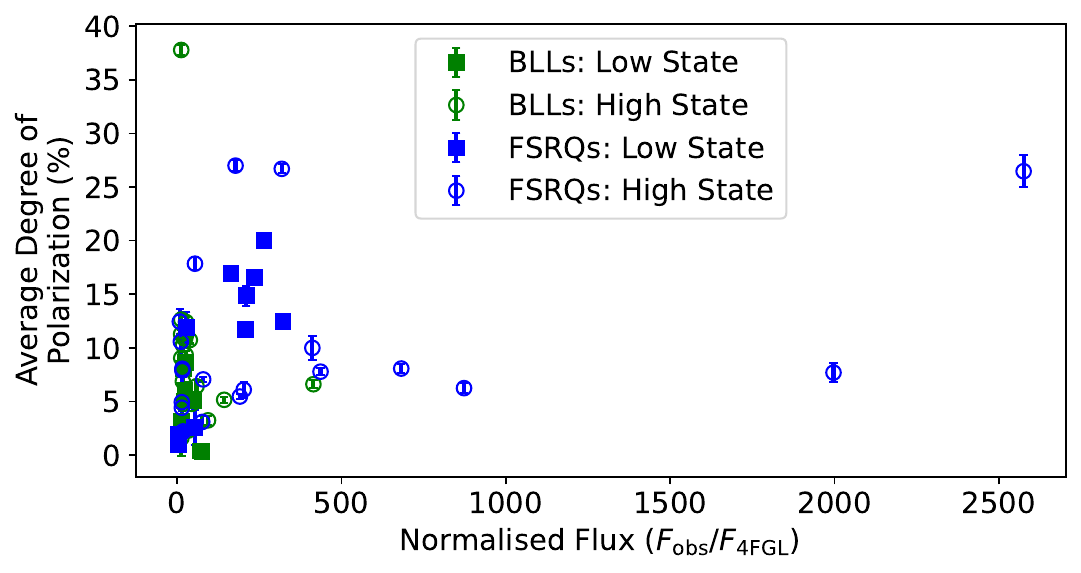}
    \caption{The relation between the averaged degree of polarization and the normalized $\gamma$-ray flux for all sources in the sample. The open circles indicate the normalized flux during high/active states, and the filled squares indicate the normalized flux during low/quiescent states. Green data points indicate values associated with BLLs, and blue data points indicate values associated with FSRQs.}
    \label{fig:Pol_vs_NormFlux_ALL}
\end{figure}

A possible correlation between the normalized $\gamma$-ray flux ($F_{\rm obs} / F_{\rm 4FGL}$, where $F_{\rm 4FGL}$ is the \textit{Fermi} 4FGL catalogue average) and the average degree of polarization was investigated (Fig.~\ref{fig:Pol_vs_NormFlux_ALL}). For the entire sample, a significant correlation of $\rho = 0.623$ (p-value = $5.223 \times 10^{-3}$) for low states was found. For high states, $\rho = -0.068$ (p-value = 0.671); thus, no correlation was found.

\begin{figure}
    \centering
    \includegraphics[width=\columnwidth]{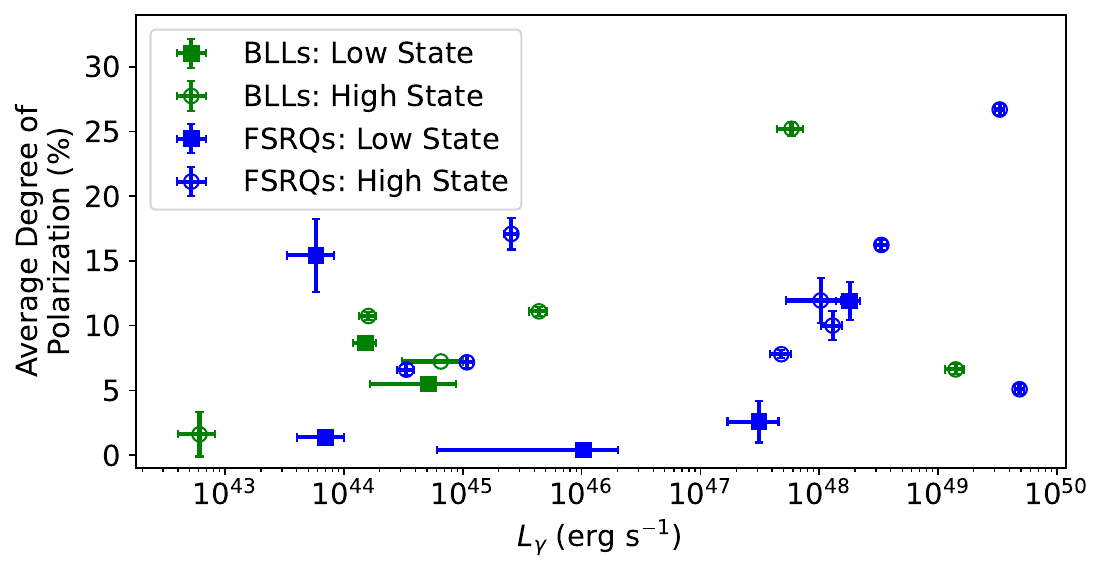}
    \caption{The averaged degree of polarization of the entire blazar sample vs. the $\gamma$-ray luminosity ($L_{\gamma}$) of each source. The open circles indicate the averaged degree of linear polarization of blazars for high/active states, and the filled squares indicate the averaged degree of linear polarization of blazars for low/quiescent states. Green data points indicate values associated with BLLs, and blue data points indicate values associated with FSRQs.}
    \label{fig:Pol_vs_Lgam}
\end{figure}

No clear correlation was found between the polarization and the $\gamma$-ray luminosity (Fig.~\ref{fig:Pol_vs_Lgam}). In low states, a Spearman's test gave $\rho = -0.179$ (p-value = 0.702), and in high states, $\rho = 0.296$ (p-value = 0.283). This erratic behaviour of the average degree of polarization with the $\gamma$-ray luminosity is in agreement with what has been found by the RoboPol campaign \citep{2016MNRAS.463.3365A}.

\begin{figure}
    \centering
    \includegraphics[width=\columnwidth]{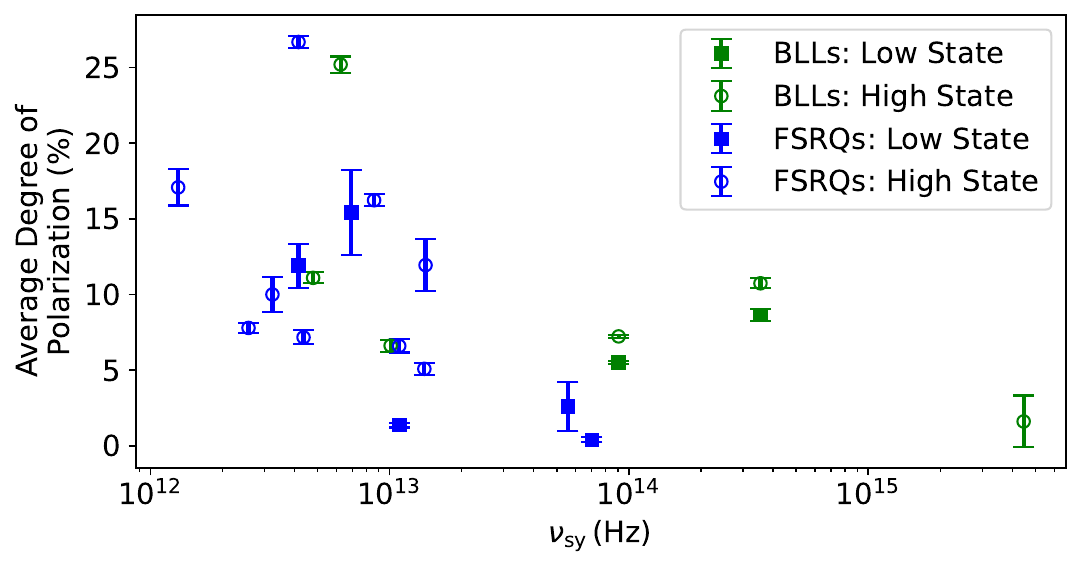}
    \caption{The averaged degree of polarization of the entire blazar sample vs. the synchrotron peak frequencies ($\nu_{\rm sy}$) of each source. The open circles indicate the averaged degree of linear polarization of blazars for high/active states, and the filled squares indicate the averaged degree of linear polarization of blazars for low/quiescent states. Green data points indicate values associated with BLLs, and blue data points indicate values associated with FSRQs.}
    \label{fig:Pol_vs_SynchFreq}
\end{figure}

\cite{2016MNRAS.463.3365A} found that, for the blazars studied in the RoboPol monitoring campaign, there appears to be an anti-correlation between the location of the synchrotron peak frequency and the average degree of polarization, with a Spearman's test resulting in $\rho = -0.300$ (p-value = $2 \times 10^{-3}$). Even though the sample studied here is much smaller, the same anti-correlation (albeit not as significant) was found for both FSRQs and BLLs, during both high an low states (see Fig.~\ref{fig:Pol_vs_SynchFreq}). For this sample of blazars, a Spearman's test yielded $\rho = -0.357$ (p-value = 0.432) in low states, and $\rho = -0.414$ (p-value = 0.125) in high states. 
The anti-correlation between the synchrotron peak frequency and the degree of polarization may hint towards a possible connection to the magnetisation of the system, or the so-called blazar sequence \citep{1998MNRAS.299..433F, 1998MNRAS.301..451G}. As was suggested by \citet{2016MNRAS.463.3365A}, the shock-in-jet model implies that the most energetic particles -- which are responsible for emitting at frequencies near the synchrotron peak frequency -- will be concentrated immediately downstream of the shock. This is the region in which the magnetic field is compressed by the shock. Thus, the further away from the shock, the less polarized the observed emission will be.
For low synchrotron-peaked FSRQs and BLLs, the synchrotron peak is situated in the infrared/optical. Therefore, the highest degrees of polarization will be observed in the optical regime. Similarly, for high synchrotron-peaked BLLs, the highest degrees of polarization will be in the X-ray regime. This then explains why the anti-correlation between the polarization and synchrotron peak frequency exists.

Lastly, no correlation was found between the observed degree of polarization and the redshift of the blazars, which is in agreement with what was found by the RoboPol monitoring campaign \citep{2016MNRAS.463.3365A}.

\subsection{The nature of polarization in blazars}
\label{subsec:Model}

Our results indicate that the polarization characteristics of blazars are not just dependent on the state of activity, i.e., a high flux state does not necessarily equate to a high degree of polarization. In fact, the polarization depends on a multitude of aspects, including the synchrotron peak frequency of the source, the ordering of the magnetic field, etc. To illustrate how the polarization is affected by the ordering of the magnetic field and the synchrotron and $\gamma$-ray flux, a model was constructed to show how the SED and frequency dependent polarization will change by varying these parameters. The SED modelling was performed using \textsc{agnpy} \citep{2022A&A...660A..18N}. A one-zone model was assumed, and the initial model parameters were based on the results of \citet{2022ApJ...925..139S} for the FSRQ 4C+01.02 (summarised in Table~\ref{tab:Model_Parameters}). Note that these parameters are not representative of all blazars, specifically not for BLLs, but used only for a demonstration of how polarization and its frequency dependence change by varying the flux and magnetic field parameters. The thermal accretion disc component was modelled using a Shakura and Sunyaev accretion disc model \citep{1973A&A....24..337S} which is, again, appropriate for an FSRQ-type blazar, but likely not for BLLs, whose accretion flows are often found to be radiatively inefficient. The synchrotron emission was calculated for a broken power-law energy distribution of relativistic electrons. The degree of synchrotron polarization is calculated as \citep[see e.g.][]{2022ApJ...925..139S},
\begin{equation}
\Pi^{\rm sy}_{\omega} = F_{\rm B} \cdot \frac{ \int N_{e}(\gamma) x(\gamma) K_{2/3} \left( x(\gamma) \right) d\gamma }{ \int N_{e}(\gamma) x(\gamma) \int^{\infty}_{x(\gamma)} K_{5/3} \left( x(\xi) \right) d\xi d\gamma },
\end{equation}
where $N_{e}$ is the electron distribution, $K_{2/3}$ and $K_{5/3}$ are Bessel functions of the second kind of order $2/3$ and $5/3$, respectively, and $F_{\rm B}$ is introduced to indicate the level of ordering of the magnetic field.

\begin{table}
\begin{center}
 \caption{The 
 Baseline parameters for the emission region, its magnetic field, the electron population, as well as some properties of the blazar's accretion disc and SMBH.}
 \label{tab:Model_Parameters}
 \begin{tabular*}{0.8\columnwidth}{@{}l@{\hspace*{14pt}}l@{\hspace*{14pt}}}
    \hline
    Property & Value \cr
    \hline
    Magnetic field strength $B$ [G] & 0.82 \cr
    Bulk Lorentz factor $\Gamma$ & 15 \cr
    Blob radius $R_{\rm blob}$ [cm] & $3 \times 10^{17}$ \cr
    Minimum energy $\gamma_{\rm min}$ & $54.8$ \cr
    Break energy $\gamma_{\rm break}$ & $7.27 \times 10^{2}$ \cr
    Maximum energy $\gamma_{\rm max}$ & $4 \times 10^{3}$ \cr
    Electron spectral index	$p_{1}$ & $2.62$ \cr
    Electron spectral index	$p_{2}$ & $3.00$ \cr
    Normalisation of electron spectrum $k$ [cm$^{-3}$] & $0.5 \times 10^{-5}$ \cr
    Black hole mass $M_{\rm SMBH}$ [$M_{\odot}$] & $3 \times 10^{9}$ \cr
    Disk luminosity $L_{\rm AD}$  [$\rm erg/s$]& $4.5 \times 10^{46}$ \cr
    Redshift $z$ & 0.35 \cr
    \hline
 \end{tabular*}
 \end{center}
\end{table}

\begin{figure}
    \centering
    \includegraphics[width=\columnwidth]{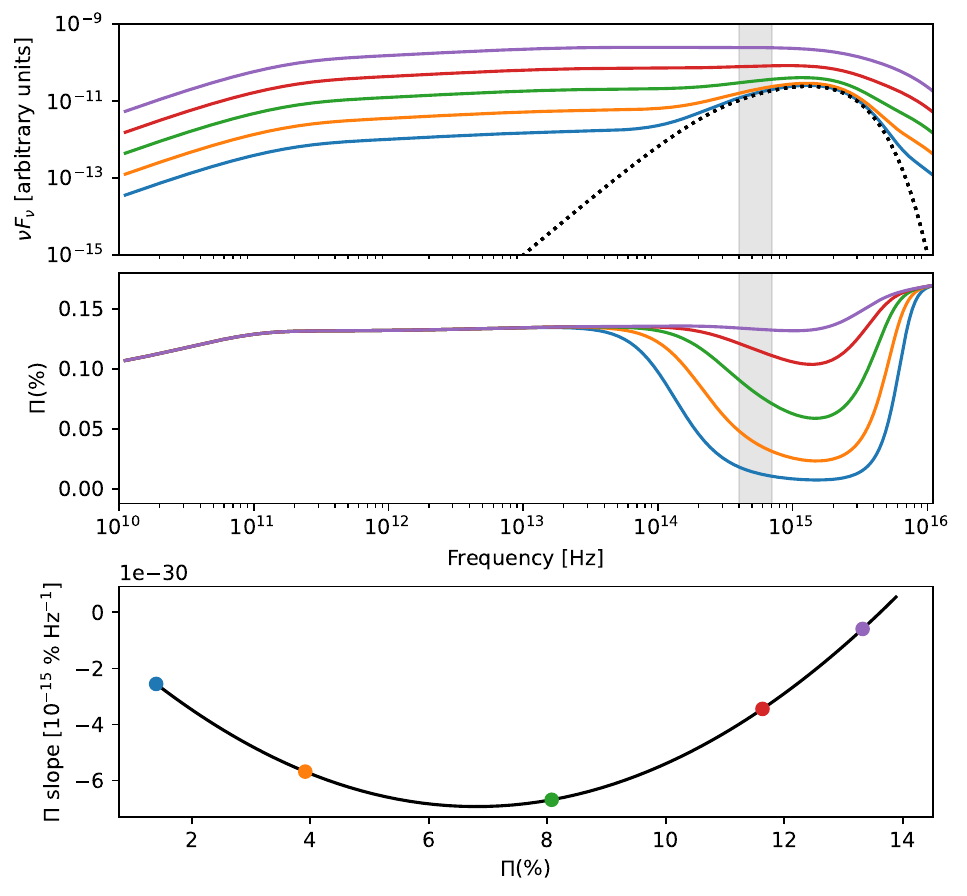}
    \caption{An illustration of the low-energy component of a blazar SED (top panel) at varying synchrotron flux levels, with the corresponding expected degree of polarization (middle panel). The solid coloured lines show the combined synchrotron plus thermal disc emission at different flux levels. The dashed black line shows the accretion disc (thermal) component. From bottom to top, the coloured lines indicate the flux as increased by an arbitrary factor of 3.5, 12.3, 43.2, and 151.7, from the lowest flux level considered (solid blue line). The grey shaded region indicates the optical frequency range. The bottom panel shows how the frequency dependence of the polarization will change with the degree of polarization at optical frequencies as the flux levels vary, with coloured marked circles corresponding to the models in the top panels.}
    \label{fig:sed_slope_flux}
\end{figure}

First, to see how an increase in the non-thermal flux will affect the frequency dependence of the polarization, the normalisation constant ($n_{0}$) of the electron distribution was increased by factors of up to $500$, with all of the other parameters fixed. The top panel of Fig.~\ref{fig:sed_slope_flux} shows the low-energy component of the SED at varying non-thermal flux levels with a fixed accretion disc component, whereas the middle panel shows the degree of polarization, with the vertical grey strip indicating the optical regime ($\nu = [4 - 7] \times 10^{14}\,{\rm Hz}$). The bottom panel shows how the frequency dependence of the polarization in the optical regime will change with increasing flux levels, first becoming increasingly negative, as the effect of the accretion disc component becomes less dominant, before becoming less negative when the non-thermal synchrotron flux becomes completely dominant over the thermal contribution.

\begin{figure}
    \centering
    \includegraphics[width=\columnwidth]{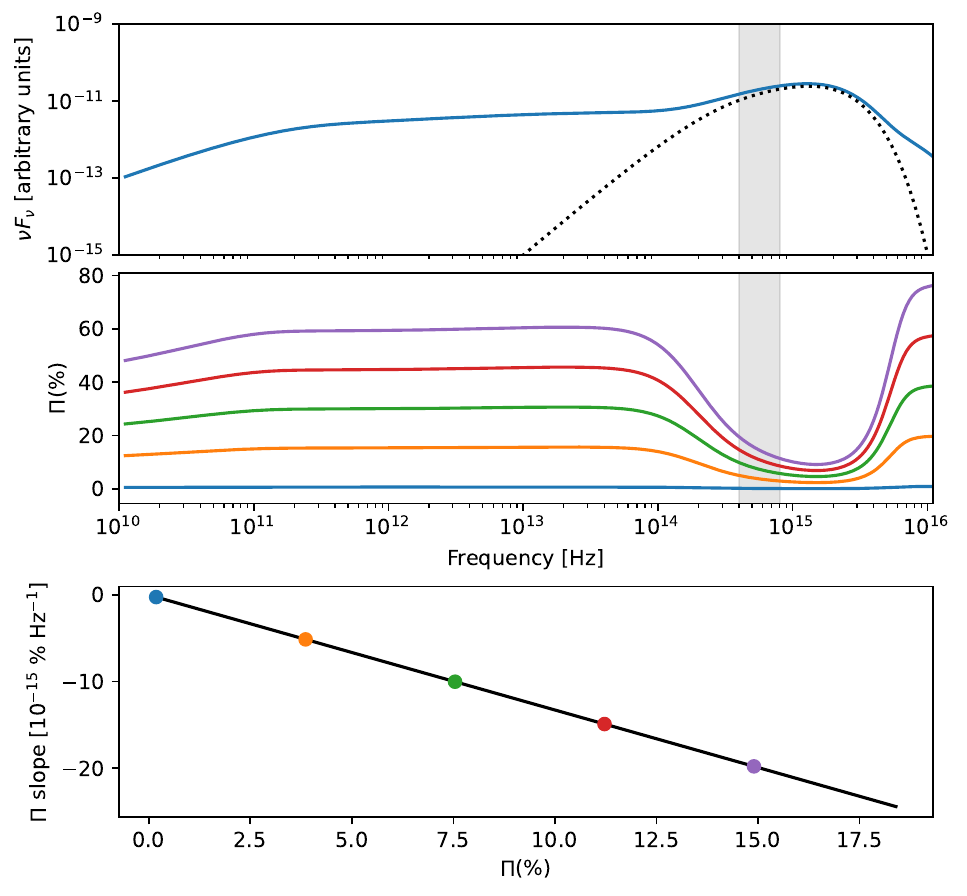}
    \caption{ Same as Fig. \ref{fig:sed_slope_flux}, but for constant synchrotron flux and varying magnetic-field ordering parameters, arbitrarily chosen as $F_{\rm B}$ = 0.01, 0.21, 0.41, 0.61, and 0.81.}
    \label{fig:sed_slope_fb}
\end{figure}

Second, the ordering of the magnetic field ($F_{\rm B}$) was increased, ranging from $F_{\rm B} = 0.01 - 1.00$, effectively increasing the degree of synchrotron polarization. The top panel of Fig.~\ref{fig:sed_slope_fb} shows the low-energy component of the SED at a constant flux level and fixed accretion disc component, whereas the middle panel shows the varying levels of polarization as the ordering of the magnetic field ($F_{\rm B}$) changes. This predicts that, as the average degree of polarization increases, the frequency dependence of the polarization will become increasingly negative in the optical regime (bottom panel). This shows that the frequency dependent polarization can be changed without changing the level of synchrotron flux.

This simple, first-approximation model shows that the most simplistic idea that increased non-thermal flux results in higher degrees of polarization and more positive frequency dependencies of the polarization, is not necessarily true and that the ordering of the magnetic field within the region where the emission is produced, also plays a critical role. More subtle effects are expected for a change in the spectral index of the relativistic electron distribution. Specifically, a steeper spectral index will increase the degree of synchrotron polarization, while potentially decreasing the optical synchrotron flux, with the two effects counter-acting each other.
While the trend of increasingly negative frequency dependencies as the degree of polarization increases is found for some individual sources (for example, 4FGL J 0231.2--4745, PKS 1510--089, and PKS 0346--279), the effect is not seen for the population as a whole (see Fig.~\ref{fig:PolSlope_vs_Pol}). This is likely due to the fact that many sources were only sampled once or twice, and not necessarily during two different states of activity. It should be noted that this model is simply a sketch of some possibilities to illustrate the complexity of the observed polarization, as it is a result of interplay between various factors. For sources like AP Lib, where the frequency dependence becomes increasingly positive as the degree of polarization increases, a more detailed model is required in which the jet's magnetic field structure is investigated. However, in order to further test the hypotheses outlined by this simplistic model, higher cadence observations are required during both high and low states of activity.

\section{Conclusions}
\label{sec:conclusions}

Investigating the spectropolarimetric behaviour of blazars is a critical step towards a better understanding of the physics underpinning the observed behaviour of blazars, as much of it remains unclear. Polarimetry is a rapidly growing field of study in understanding high energy emission from blazars, and could potentially provide the necessary information to fully understand blazar emission and constrain the emission mechanisms producing it. To this end, an optical spectropolarimetry campaign of a number of high-energy blazars was undertaken. This study, while significantly smaller than the programmes undertaken by RoboPol and the Steward observatory, provided valuable insights into the level and frequency dependence of the optical polarization 
of high-energy blazars.

Two blazars exhibiting unusual behaviour were observed, namely AP Lib and PKS 1510--089. The $\gamma$-ray and optical emission of AP Lib showed large-scale, long-term variability during the first half of 2021, with changes in the degree of linear polarization lagging behind the behaviour of the light curves by $\sim56\,$days. This indicated a possible large-scale change in the structure of the jet. For PKS 1510--089, the jet-component appeared to have ``switched off'' entirely in July 2021, with no intrinsic polarization observed in the optical spectropolarimetric data. In addition, the frequency dependence of the polarization in 2021 (prior to the flux-drop) indicated a strong anti-correlation, becoming increasingly negative with increasing degree of polarization (as shown in Fig.~\ref{fig:PKS1510_PolSlope_vs_Pol}). Though we caution the correlation found for PKS 1510--089 during 2021 (as discussed in Section~\ref{subsubsec:PKS1510}), this is the opposite of what has been observed for AP Lib and PKS 1034--293.

The lack of strong correlations between the polarization and other emission features for the sources in this sample may be due to the uneven sampling of sources. Stronger correlations were found for individual sources, and further investigation is warranted. However, the overall results from this study are in general agreement with the findings from the RoboPol monitoring campaign, despite the smaller sample of sources and the bias towards flaring activity.

The results in this study highlight that modelling of the nature of polarization and its frequency dependence is required to fully understand these highly energetic sources. 
Detailed multi-wavelength modelling of some individual sources would also be required to explain some of the more extreme behaviours observed in this sample and will be subject of future work. This will provide an objective study of the optical spectropolarimetric behaviour of blazars, and will serve as an excellent complement to the predicted behaviour of the polarization in the X-ray regime.

\section*{Acknowledgements}

J.B. acknowledges support by the National Research Foundation of South Africa (NRF, grant number MND210616612361). Opinions expressed and conclusions arrived at, are those of the author and are not
necessarily to be attributed to the NRF. 
This work was supported by the Department of Science and Technology and the
National Research Foundation of South Africa through a block grant
to the South African Gamma-Ray Astronomy Consortium.
BvS acknowledges support by the National Research Foundation of South Africa (grant number 119430).
M.Z. acknowledges funding by the Deutsche Forschungsgemeinschaft (DFG, German Research Foundation) -- project number 460248186 (PUNCH4NFDI). 
The South African Astronomical Observatory: Some of the observations reported in this paper were
obtained with the Southern African Large Telescope (SALT) under program 2021-2-LSP-001 (PI:
D.A.H. Buckley) and program 2019-2-MLT-001 (PI: B. van Soelen).
This research has made use of the SIMBAD database, operated at CDS, Strasbourg, France, as well as the NASA/IPAC Extragalactic Database (NED), which is funded by the National Aeronautics and Space Administration and operated by the California Institute of Technology.
This research has made use of the VizieR catalogue access tool, CDS, Strasbourg, France.

\section*{Data Availability}

The optical data is available upon reasonable request from the authors. The \textit{Fermi}-LAT data is publicly available.



\bibliographystyle{mnras}
\bibliography{MSc_bib}



\appendix

\section{Summary of observational results}
\label{app:full_results}

Table~\ref{tab:full_results} provides a summary of all the observational results obtained for the entire sample of blazars investigated in this project. This includes the $\gamma$-ray state (i.e. active or quiescent), whether the fluxes were rising or falling, the average degree of polarization and polarization angle, the frequency dependence of the polarization -- where a positive ($+$) slope indicates an increase in polarization towards higher frequencies, and a negative ($-$) slope indicates a decrease in polarization towards higher frequencies -- and the isotropic-equivalent $\gamma$-ray luminosity on the respective dates of observation. The $\gamma$-ray states were defined as active/high based on what has been reported in literature, and/or when the observed fluxes were greater than the daily averaged flux in the \textit{Fermi} 4FGL catalogue by a factor of $\sim10$.\\

\begin{landscape}
 \begin{table}
 \begin{center}
  \caption{A summary of the SALT observational results, with the $\gamma$-ray state, and whether it was rising or falling, the average degree of polarization ($\langle\Pi\rangle$) and average polarization angle ($\langle\rm{PA}\rangle$), taken between $\lambda = 4100$\,\rm{\AA} and $6200$\,\rm{\AA} to obtain a consistent result for all of the sources across the various instrumental setups used. The $\gamma$-ray luminosity is obtained from the average flux values for each source on the dates of observation. The luminosity distance $D_L$ was calculated from the redshifts of each target, and a cosmology was adopted with $H_0 = 70\,\rm{km\,s^{-1}\,Mpc^{-1}}$ and $\Omega_{m} = 0.3$. The frequency dependence of the degree of polarization ($m_{\Pi}$) is indicated by either a $+$ or $-$, indicating whether the degree of polarization increases or decreases towards higher frequencies, respectively. Note that, where empty, the frequency dependence of the degree of polarization was flat and/or comparable to that of the comparison star.}
  \label{tab:full_results}
  \begin{tabular}{ld{4.1}ccc*{2}{S}cc}
    \hline
    Target & \multicolumn{1}{c}{$D_L$ ($\rm{Mpc}$)} & Obs. Date & Gamma-ray State & Rise/Fall & \multicolumn{1}{c}{$\langle\Pi\rangle$ ($\%$)} & \multicolumn{1}{c}{$\langle\rm{PA}\rangle$ ($^{\circ}$)} & $m_{\Pi}$ & \multicolumn{1}{c}{$L_\gamma$ ($\rm{erg\,s^{-1}}$)} \\
    \hline
    PKS 0426--380 & 7525.73  & 2017-01-21 & Active & Fall & 10.9(0.3) & 11.0(1.0) & + & ($5.5\pm0.8)\times10^{45}$ \cr
                  &            & 2017-02-20 & Active & Fall & 11.3(0.4) & 50.2(0.9) & + & ($3.3\pm0.7)\times10^{45}$ \cr
    PKS 0447--439 & 494.82   & 2017-02-21 & Active & Fall & 1.6(1.7) & 151.1(45.5) & + & ($6.1\pm2.1)\times10^{42}$ \cr
    TXS 0506+056  & 1776.71   & 2017-10-15 & Active & Fall & 10.7(0.3) & 29.8(0.8) & + & ($1.6\pm0.3)\times10^{44}$ \cr
                  &             & 2017-10-21 & Quiescent &  & 8.7(0.4) & 52.6(1.0) & -- & ($1.5\pm0.3)\times10^{44}$ \cr
    PKS 0537--441 & 5737.33  & 2019-01-14 & Active & Rise & 37.8(0.5) & 9.6(0.2) & + & ($3.9\pm1.1)\times10^{47}$ \cr
                  &             & 2019-03-05 & Active & Fall & 12.7(0.6) & 8.5(1.0) & + & ($7.8\pm1.6)\times10^{47}$ \cr
    PKS 1454--354 & 10229.91 & 2021-06-13 & Active & Fall & 6.6(0.4) & 123.2(1.4) &   & ($1.4\pm0.3)\times10^{49}$ \cr
    AP Lib        & 217.69    & 2020-05-14 & Quiescent &  & 5.0(0.4) & 147.0(3.1) & + & ($6.6\pm5.3)\times10^{44}$  \cr
                  &             & 2020-06-08 & Quiescent &  & 5.1(0.6) & 145.2(2.6) & + & ($6.1\pm4.5)\times10^{44}$ \cr
                  &             & 2020-06-24 & Quiescent &  & 3.2(0.6) & 165.9(3.8) & + & ($4.0\pm3.1)\times10^{44}$ \cr
                  &          & 2020-08-07 & Quiescent &  & 8.1(0.3) & 126.7(1.6) & + & ($4.9\pm2.4)\times10^{44}$ \cr
                  &          & 2020-09-04 & Quiescent &  & 6.1(0.7) & 132.3(2.8) & + & ($4.3\pm2.3)\times10^{44}$ \cr
                  &          & 2021-02-18 & Active    & Rise & 2.3(0.5) & 114.6(3.9) & -- & ($6.2\pm1.9)\times10^{44}$ \cr
                  &          & 2021-03-12 & Active    & Rise & 4.8(0.3) & 130.0(1.5) & + & ($6.7\pm2.2)\times10^{44}$ \cr
                  &          & 2021-04-04 & Active    & Fall & 8.3(0.4) & 130.5(1.3) & + & ($6.2\pm3.3)\times10^{44}$ \cr
                  &          & 2021-04-11 & Active    & Fall & 9.4(0.4) & 124.9(0.8) & + & ($6.9\pm3.7)\times10^{44}$ \cr
                  &          & 2021-04-21 & Active    & Fall & 10.9(0.5) & 132.9(0.9) & + & ($6.5\pm3.2)\times10^{44}$ \cr
                  &          & 2021-05-02 & Active    & Fall & 9.1(0.6) & 143.5(1.4) & + & ($7.3\pm3.6)\times10^{44}$ \cr
                  &          & 2021-05-08 & Active    & Fall & 10.3(0.6) & 148.5(1.1) & + & ($6.3\pm4.1)\times10^{44}$ \cr
                  &          & 2021-05-14 & Active    & Fall & 6.9(0.6) & 145.4(1.7) & + & ($6.7\pm4.1)\times10^{44}$ \cr
                  &          & 2021-06-03 & Active    & Rise & 6.4(0.6) & 156.8(0.9) & + & ($7.4\pm4.3)\times10^{44}$ \cr
                  &          & 2021-06-09 & Active    & Rise & 3.3(0.4) & 143.2(4.0) & + & ($5.7\pm3.7)\times10^{44}$ \cr
                  &          & 2021-06-10 & Active    & Rise & 5.2(0.3) & 145.1(1.1) & + & ($6.2\pm4.4)\times10^{44}$ \cr
                  &          & 2021-07-10 & Active    & Fall & 4.8(0.4) & 146.0(1.8) & + & ($6.0\pm3.6)\times10^{44}$ \cr
                  &          & 2021-07-31 & Active    & Rise & 12.4(0.5) & 142.0(0.7) & + & ($7.0\pm2.8)\times10^{44}$ \cr
    PKS 0035--252 & 2819.22 & 2018-07-21 & Active & Fall & 7.8(0.3) & 78.7(1.3) & -- & ($4.8\pm1.0)\times10^{47}$ \cr
    PKS 0131--522 & 86.97    & 2017-11-19 & Active & Rise & 8.1(0.4) & 2.8(1.9) & -- & ($9.3\pm1.3)\times10^{44}$ \cr
                  &          & 2017-11-22 & Active & Fall & 6.3(0.5) & 7.5(2.3) & -- & ($1.2\pm0.2)\times10^{45}$ \cr
    PKS 0208--512 & 6631.56 & 2019-12-05 & Active & Rise & 27.0(0.5) & 20.3(0.9) & -- & ($3.2\pm0.4)\times10^{48}$ \cr
                  &          & 2019-12-19 & Active & Fall & 5.5(0.3) & 97.0(1.7) & + & ($3.5\pm0.3)\times10^{48}$ \cr
    4FGL J0231.2--4745 & 4749.09 & 2019-10-23 & Active & Rise & 26.5(1.5) & 144.0(1.1) & -- & ($2.8\pm0.3)\times10^{45}$ \cr
                  &          & 2019-10-29 & Active & Fall & 7.7(0.9) & 132.9(3.2) & -- & ($2.3\pm0.3)\times10^{45}$ \cr
    PKS 0346--279 & 6533.63 & 2018-02-09 & Active & Rise & 17.9(0.6) & 116.3(1.0) & -- & ($1.0\pm0.5)\times10^{48}$ \cr
                  &          & 2021-11-05 & Active & Rise & 6.1(0.8) & 79.2(2.7) & + & ($3.3\pm0.4)\times10^{48}$ \cr
    PKS 0837+012  & 7626.43  & 2021-03-16 & Active & Fall & 10.0(1.2) & 132.5(3.1) & -- & ($1.3\pm0.3)\times10^{48}$ \cr
    PKS 0907-023  & 6257.77  & 2017-01-19 & Quiescent &  & 2.6(1.6) & 58.8(26.0) & -- & ($3.1\pm1.4)\times10^{47}$ \cr
    PKS 1034--293 & 1612.60  & 2020-05-15 & Quiescent &  & 16.6(0.5) & 105.5(0.7) & + & ($5.8\pm2.5)\times10^{43}$ \cr
                  &          & 2021-02-08 & Quiescent &  & 17.0(0.7) & 98.1(1.2) & + & ($4.1\pm2.4)\times10^{43}$ \cr
                  &          & 2021-03-17 & Quiescent &  & 14.9(0.9) & 13.6(1.8) & + & ($5.2\pm2.5)\times10^{43}$ \cr
    \hline
  \end{tabular}
  \end{center}
 \end{table}
\end{landscape}

\begin{landscape}
 \begin{table}
 \begin{center}
  \contcaption{}
  \label{tab:full_results_cont}
  \begin{tabular}{ld{4.1}ccc*{2}{S}cc}
    \hline
    Target & \multicolumn{1}{c}{$D_L$ ($\rm{Mpc}$)} & Obs. Date & Gamma-ray State & Rise/Fall & \multicolumn{1}{c}{$\langle\Pi\rangle$ ($\%$)} & \multicolumn{1}{c}{$\langle\rm{PA}\rangle$ ($^{\circ}$)} & $m_{\Pi}$ & \multicolumn{1}{c}{$L_\gamma$ ($\rm{erg\,s^{-1}}$)} \\
    \hline
                       &         & 2021-04-09 & Quiescent &  & 12.4(0.6) & 161.2(1.2) & + & ($8.0\pm2.5)\times10^{43}$ \cr
                   &          & 2021-06-05 & Quiescent &  & 20.0(0.7) & 135.7(1.0) & + & ($6.6\pm2.4)\times10^{43}$ \cr
                   &          & 2021-06-11 & Quiescent &  & 11.7(0.6) & 132.3(1.5) & + & ($5.2\pm2.4)\times10^{43}$ \cr
3C 273       & 754.94 & 2017-06-13 & Quiescent &  & 0.4(0.2) & 27.2(22.2) &  & ($9.9\pm2.2)\times10^{45}$ \cr
             &        & 2017-06-14 & Quiescent &  & 0.4(0.2) & 43.3(19.3) &  & ($1.1\pm0.2)\times10^{46}$ \cr
PKS 1424--418      & 11105.97 & 2022-07-26 & Active & Rise & 7.1(0.3) & 160.4(3.0) &  & ($4.9\pm0.4)\times10^{49}$ \cr
                   &          & 2022-08-15 & Active & Fall & 3.1(0.5) & 141.1(3.7) & + & ($4.8\pm0.3)\times10^{49}$ \cr
PKS 1510--089      & 1919.07 & 2021-04-06 & Active & Fall & 7.9(1.4) & 13.0(2.2) & -- & ($1.2\pm0.3)\times10^{47}$ \cr
                   &          & 2021-04-09 & Active & Fall & 5.0(1.2) & 4.7(2.7) & -- & ($1.1\pm0.2)\times10^{47}$ \cr
                   &          & 2021-04-17 & Active & Fall & 2.2(0.4) & 100.3(4.5) & -- & ($1.4\pm0.3)\times10^{47}$ \cr
                   &          & 2021-04-20 & Active & Fall & 2.2(0.5) & 178.9(4.8) & -- & ($1.4\pm0.2)\times10^{47}$ \cr
                   &          & 2021-05-08 & Active & Rise & 12.5(1.1) & 155.2(0.8) & -- & ($6.3\pm2.0)\times10^{46}$ \cr
                   &          & 2021-05-13 & Active & Rise & 10.6(1.6) & 31.4(1.5) & -- & ($9.3\pm2.5)\times10^{46}$ \cr
                   &          & 2021-06-05 & Active & Fall & 4.4(0.3) & 146.6(2.2) & -- & ($1.1\pm0.3)\times10^{47}$ \cr
                   &          & 2021-06-10 & Active & Fall & 8.1(0.8) & 81.2(1.3) & -- & ($1.3\pm0.5)\times10^{47}$ \cr
                   &          & 2022-04-25 & Quiescent &  & 1.5(0.7) & 57.0(22.5) &  & ($2.9\pm3.5)\times10^{46}$ \cr
                   &          & 2022-04-26 & Quiescent &  & 1.2(0.6) & 57.6(23.8) &  & ($2.9\pm3.5)\times10^{46}$ \cr
                 &          & 2022-05-24 & Quiescent &  & 1.7(0.7) & 58.9(11.9) &  & ($3.7\pm2.7)\times10^{46}$ \cr
                 &          & 2022-05-25 & Quiescent &  & 1.1(0.4) & 74.5(11.4) &  & ($3.7\pm2.7)\times10^{46}$ \cr
                 &          & 2022-05-30 & Quiescent &  & 2.0(1.0) & 49.0(13.9) &  & ($3.6\pm3.6)\times10^{46}$ \cr
                 &          & 2022-06-05 & Quiescent &  & 1.2(0.4) & 71.6(11.7) &  & ($4.3\pm3.2)\times10^{46}$ \cr
                 &          & 2022-06-20 & Quiescent &  & 1.2(0.5) & 71.3(18.3) &  & ($3.3\pm3.1)\times10^{46}$ \cr
                 &          & 2022-06-26 & Quiescent &  & 1.0(0.5) & 83.9(22.9) &  & ($3.3\pm3.2)\times10^{46}$ \cr
                 &          & 2022-07-27 & Quiescent &  & 1.2(0.5) & 63.4(27.6) &  & ($2.7\pm2.4)\times10^{46}$ \cr
                 &          & 2022-07-28 & Quiescent &  & 1.5(0.8) & 77.6(21.4) &  & ($3.6\pm3.1)\times10^{46}$ \cr
                 &          & 2022-07-31 & Quiescent &  & 1.4(0.5) & 63.7(14.4) &  & ($3.8\pm3.2)\times10^{46}$ \cr
PKS 2023--07     & 9911.33 & 2016-04-16 & Active & Rise & 26.7(0.4) & 167.3(0.7) & -- & ($3.3\pm0.3)\times10^{49}$ \cr
                 &          & 2018-10-04 & Quiescent &  & 11.9(1.5) & 88.0(2.7) & + & ($1.8\pm0.4)\times10^{48}$ \cr
    \hline
  \end{tabular}
  \end{center}
 \end{table}
\end{landscape}


\section{A note on the effects of interstellar polarization}
\label{app:ISP_info}

Assuming the relations given in Equations \ref{eqn:ISP_max} - \ref{eqn:DeltaPA_max}, Table \ref{tab:ISP_properties} provides the interstellar extinction along each target's line of sight, along with the typical and maximum expected ISP values, as well as the median value of the maximum polarization angle change for each source.

\begin{table*}
\begin{center}
\small
\caption{A summary of the Galactic extinction \citep{2011ApJ...737..103S} along each target's line of sight, along with the corresponding typical and maximum ISP levels \citep{1975ApJ...196..261S, Smith04}, and the maximum polarization angle change that can be expected due to the effects of the ISP \citep{2005MNRAS.363.1241H} .}
\label{tab:ISP_properties}
\begin{tabular}{lcccc}
\hline
Target & Galactic Extinction E(B-V) & ISP$_{\rm{typical}}$ $(\%)$ & ISP$_{\rm{max}}$ $(\%)$ & $\Delta \rm{PA}_{\rm{max}}$ ($^\circ$) \\
\hline
PKS 0426--380      & 0.021 & 0.062 & 0.18 & 0.48 \cr
PKS 0447--439      & 0.012 & 0.035 & 0.10 & 1.85 \cr 
TXS 0506+056       & 0.092 & 0.28  & 0.83 & 2.49 \cr
PKS 0537--441      & 0.032 & 0.095 & 0.28 & 0.43 \cr
PKS 1454--354      & 0.086 & 0.26  & 0.78 & 3.37 \cr
AP Lib             & 0.12  & 0.35  & 1.06 & 4.88 \cr
PKS 0035--252      & 0.013 & 0.039 & 0.17 & 0.43 \cr
PKS 0131--522      & 0.02  & 0.064 & 0.19 & 0.78 \cr
PKS 0208--512      & 0.017 & 0.052 & 0.17 & 0.49 \cr
4FGL J0231.2--4745 & 0.013 & 0.039 & 0.12 & 0.28 \cr
PKS 0346--279      & 0.0094 & 0.028 & 0.085 & 0.27 \cr
PKS 0837+012       & 0.041 & 0.12  & 0.36 & 1.04 \cr
PKS 0907--023      & 0.022 & 0.067 & 0.20 & 2.22 \cr
PKS 1034--293      & 0.044 & 0.13  & 0.40 & 0.73 \cr
3C 273             & 0.018 & 0.054 & 0.16 & 12.30 \cr
PKS 1424--418      & 0.11 & 0.32   & 0.95 & 6.38 \cr
PKS 1510--089      & 0.085 & 0.26  & 0.77 & 13.48 \cr
PKS 2023--07       & 0.033 & 0.098 & 0.30 & 0.51 \cr
\hline
\end{tabular}
\end{center}
\end{table*}

\bsp	
\label{lastpage}
\end{document}